\documentclass[11pt]{amsart}

\usepackage[utf8]{inputenc}
\usepackage[T1]{fontenc}
\usepackage{lmodern}
\usepackage{fullpage} 

\usepackage{amsmath,appendix}
\usepackage{physics}

\usepackage{pdflscape}
 
\usepackage[normalem]{ulem}
\usepackage{hyperref} 
\hypersetup{
colorlinks   = true,
citecolor    = blue
}

\usepackage{setspace}
\theoremstyle{plain}
\newtheorem{Lem}{Lemma}
\newcommand{\cm}[2]{\bqty{#1,#2}} 
\newcommand{\qn}[1]{\bqty{#1}_q} 
\newcommand{\proj}[2]{P^{#1}_{#2}} 

\newcommand{\khalf}{\mbox{\footnotesize $\frac{n}{2}$} }

\newcommand{\re}{\mathop{\rm Re}}
\newcommand{\im}{\mathop{\rm Im}}

\begin{document}

\title[On the quantum algebra $su_q(1,1)$ from a Special Function standpoint]{On the quantum algebra $su_q(1,1)$ from a Special Function standpoint}
\author[R. \'Alvarez-Nodarse]{Renato \'Alvarez-Nodarse}
\address{Departamento de An\'alisis Matem\'atico, Universidad de Sevilla, Av.~Reina Mercedes s/n, E-41012 Sevilla, Spain}
\email{ran@us.es}
\author[A. Arenas-G\'omez]{Alberto Arenas-G\'omez}
\address{Departamento de Matem\'aticas y Computaci\'on, Universidad de La Rioja, Calle Madre de Dios 53, 26006 Logro\~no, Spain}
\email{alberto.arenas@unirioja.es}
\date{\today}

\maketitle


\begin{abstract} 
In this paper, we study the tensor product of two unitary irreducible representations, as well as the tensor product of a unitary irreducible representation with a finite-dimensional one, and determine the corresponding Clebsch–Gordan coefficients. Using von Neumann’s projection operator method, we obtain an explicit representation of these coefficients, which allows us to express them as symmetric $q$-hypergeometric series. Finally, by leveraging the properties of the $q$-hypergeometric function, we derive several properties of the Clebsch-Gordan coefficients, including a number of new results, in a unified and straightforward manner.
\end{abstract}

\section{Introduction}

It is well known that the representation theory of groups and algebras  plays a fundamental role in Physics and, in particular, in Quantum Mechanics (for a recent review see e.g.~\cite{woi:2017} and references therein). At the end of the 20th century, motivated by various problems arising in the development of the quantum inverse problem method and the Yang-Baxter equation the $q$-deformation of the classical Lie groups, the known as \textit{quantum groups} were introduced (for an historical overview, see e.g.~\cite{res:90,fad:2006}). This new topic immediately attracted the attention of the mathematical-physics community, especially after the appearance of the first models of $q$-deformed oscillators \cite{bie,mac} in 1989 based on the quantum algebra $su_q(2)$. One of the reasons for this interest was the expectation that the introduction of an additional parameter $q$ can allow a better description of some natural phenomena. For a review up to the year 2012 see~\cite[Chapter 12]{ernst}. 

In this context, it is worth mentioning that one of the earliest applications was the construction of the $q$-analog of the Quantum Theory of Angular Momenta based on the quantum algebra $su_q(2)$. For a recent review on this topic by means of the Special Function Theory see the recent work \cite{ran-aag}. At this point, we should also mention that, apart from the already referenced $q$-oscillators introduced in \cite{bie,mac}, as already mentioned based on the quantum algebra $su_q(2)$, there are several other models of $q$-deformed oscillators that are related to the quantum algebra $su_q(1,1)$; see e.g.~\cite{ran:2005,ata:2006,jaf:2012,kli:2006} and references therein. Let us point out that both algebras $su_q(2)$ and $su_q(1,1)$ are special cases of $sl_q(2)$, but they are quite different: the former is compact whereas the latter is non-compact. In fact, the unitary and irreducible representations for the $su_q(2)$ are finite-dimensional whereas the corresponding for the $su_q(1,1)$ are infinite-dimensional (see e.g.~\cite[Vol III]{vil:92}). 

In this paper, we focus on the Clebsh-Gordan coefficients of the quantum algebra $su_q(1,1)$. Our aim is to study the tensor product of two unitary and irreducible representations, and the tensor product of one unitary and irreducible representation and one finite-dimensional (and, therefore, not irreducible), and find the corresponding Clebsh-Gordan coefficients. We will use, as it was done in~\cite{ran-aag}, an approach based on the projection operators method of von Neumann. For the details on this method see \cite{tol:01,tol:05} and references therein, while for the representation theory of the quantum algebras $su_q(2)$ and $su_q(1,1)$ see e.g.~\cite{koe:2001,smi-g,vil:92}.

The explicit expression for the Clebsh-Gordan coefficients related to the tensor product of two unitary and irreducible representations of the quantum algebra $su_q(1,1)$ by means of the projection operator method was first obtained in the Master Thesis of A.~del Sol (see also \cite{smi-sol2} and \cite[\S5.4]{gromov}, where some partial results were published). The case of the tensor product of an unitary and irreducible representation and a finite-dimensional representation was studied by the first author in his Master Thesis \cite{ran-mt} (see also \cite[\S5.4]{gromov} for some partial results). Since these works are very difficult to obtain, we include here the calculation of the Clebsch-Gordan coefficients in both situations and we correct several typographical errors in the original derivations. Let us also mention that the tensor product of representations for the $su_q(1,1)$ was also studied in \cite{aiz} by using a completely different method.  

Here, we will go further in the study of the Clebsch-Gordan coefficients by also exploring the connection between them and the symmetric $q$-hypergeometric function, which was introduced by Nikiforov and Uvarov \cite[p.~138]{nsu}, in the same fashion as we did in our recent work \cite{ran-aag}. In fact, using some properties of the $q$-hypergeometric function, many properties of the Clebsch-Godan coefficients can be obtained in a straightforward way. As examples of this, we will derive its symmetry property, which requires a more elaborate proof by other methods, as well as their explicit expression in the most relevant cases. It is worth pointing out the connection between the quantum algebras and the $q$-special functions is well documented in the literature (see e.g.~\cite{koe:96,koe:2001,vil:92}).

The structure of the paper is as follows. In Section 2, we introduce the notation and some necessary preliminary results. In Section 3, we introduce the quantum algebra $su_q(1,1)$, discuss some of its main properties, and define the projection operators. In Section 4, we discuss the Clebsch-Gordan coefficients corresponding to the tensor product of two unitary and irreducible representations (positive series). Finally, in Section 5, we study the Clebsch-Gordan coefficients corresponding to the tensor product of an unitary and irreducible and a finite-dimensional (non-irreducible) representation of the quantum algebra~$su_q(1,1)$. The paper concludes with three appendices, where a sketch of the construction of the irreducible and unitary representations of $su_q(1,1)$ is presented, basic facts regarding the $q$-Hahn polynomials are exhibited, and some recurrence relations of the Clebsh-Gordan coefficients are resumed, respectively.


\section{Some preliminary results}

We will use the symmetric notation for $q$-hypergeometric series introduced by Nikiforov and Uvarov (see e.g.~\cite{ran-aag,nsu}). Let be $q\in(0,1)$ and $x\in\mathbb{R}$, then, the symmetric quantum number~$[x]_q$ is given by
\begin{equation*}
[x]_q=\frac{q^{x}-q^{-x}}{q-q^{-1}}, 
\end{equation*}
while the symmetric $q$-Pochhammer is defined for all $a\in\mathbb{R}$ by
\begin{equation*}
(a|q)_0=1,\quad (a|q)_n = \prod_{m=0}^{n-1} [a+m]_q,\quad n=1,2,3,\ldots
\end{equation*}
If $a\in\mathbb{N}$, the following relations hold
\begin{equation} \label{q-fac-poc}
(a|q)_n=\frac{[a-1+n]!}{[a-1]!}, \qquad 
(-a|q)_n=\frac{(-1)^n \, [a]!}{[a-n]!}, \quad \text{if $a\geq n$ and 0 if $a<n$},
\end{equation}
where $[n]_q!$ denotes the symmetric $q$-factorial 
\begin{equation*}
[n]_q!=(1|q)_n= [1]_q[2]_q\cdots [n-1]_q[n]_q .
\end{equation*}
 
Additionally, the symmetric $q$-hypergeometric function, which was introduced in \cite[p.~138]{nsu}, is defined by
\begin{equation}\label{q-hip-def}
  {}_{p+1}F_p
\left(\!\!\begin{array}{c} {a_1,\dots,a_{p+1}} \\ {b_1,\dots,b_p} \end{array} \,
\bigg|\, q\,,\,  z \right)\!=\!
\displaystyle \sum _{k=0}^{\infty}\frac{ (a_1|q)_k(a_2|q)_k \cdot \cdot \cdot  (a_{p+1}|q)_k}
{(b_1|q)_k(b_2|q)_k \cdot \cdot \cdot  (b_p|q)_k}\frac{z^k}{(1|q)_k}
\end{equation}
and constitutes one of the possible $q$-analogues of the generalized hypergeometric function (see e.g.~\cite{sla}). Another $q$-analogue is the so called basic hypergeometric series defined in \cite{gas}; however, for our purpose the former works better than the latter. For further discussion see~\cite{ran-aag}.

We shall restrict the present work to the case in which at least one of the numbers $a_i$, where $i=1,2,\dots,p+1$, is a nonpositive integer, which implies the terminating character of the series~\eqref{q-hip-def}. The following transformation formula
\begin{equation}\label{142q}
{}_{3}F_2\left(\!\!\begin{array}{c} {-n,a,b} \\ {d, e} \end{array} \bigg| q,  q^{\pm(a+b-n-d-e+1)} \right)\!=
\frac{q^{\pm an}(e-a|q)_n}{(e|q)_n}  
{}_{3}F_2\left(\!\!\begin{array}{c} {-n,a,d-b} \\ {d,a-e-n+1} \end{array}\bigg|q, q^{\pm(b-e)} \right),
\end{equation}
is deduced from~\cite[(III.11), p.~330]{gas}.

We also introduce some additional summation formulae that follows from the $q$-analogue of the Chu-Vandermonde formula (see e.g. \cite{ran-aag}).
In the following let us assume $n,b,c\in\mathbb{Z}^+$ such that $n<\min\{b,c\}$, then
\begin{equation*}
{}_{2}F_1\left(\!\!\begin{array}{c} {-n,-b} \\ {-c} \end{array} \bigg| q,  q^{\pm(b-c+n-1)} \right)=
\begin{cases}
\dfrac{[c-n]_q![c-b]_q!}{[c]_q![c-b-n]_q!}q^{\pm bn}, & b<c, \\[5mm]
\dfrac{(-1)^n\qn{c-n}!\qn{b+n-c-1}!}{\qn{c}!\qn{b-c-1}!}q^{\pm bn}, & b>c.
\end{cases}
\end{equation*}
The latter turns into
\begin{equation}\label{suma2equiv}
\sum_{r=0}^{n}\frac{(-1)^{r}\qn{c-r}!}{\qn{r}!\qn{b-r}!\qn{n-r}!}q^{\pm(b-c+n-1)r}=
\begin{cases}
\dfrac{[c-n]_q![c-b]_q!}{[n]_q![b]_q![c-b-n]_q!}q^{\pm bn}, & b<c, \\[5mm]
\dfrac{(-1)^n\qn{c-n}!\qn{b+n-c-1}!}{\qn{b}!\qn{n}!\qn{b-c-1}!}q^{\pm bn},& b>c,
\end{cases}
\end{equation}
by means of \eqref{q-fac-poc}. In addition,  the summation formula \begin{equation}\label{suma3}
{}_{2}F_1\left(\!\!\begin{array}{c} {-n,b} \\ {-c} \end{array} \bigg| q,  q^{\pm(b+c-n+1)} \right)  =
\dfrac{\qn{c-n}!\qn{b+c}!}{\qn{c}!\qn{b+c-n}!}q^{\pm bn}
\end{equation}
leads to
\begin{equation}\label{suma3equiv}
\sum_{r=0}^{\infty}\frac{\qn{c-r}!\qn{b+r-1}!}{\qn{r}!\qn{n-r}!}q^{\pm(b+c-n+1)r}
=\frac{\qn{c+b}!\qn{c-n}!\qn{b-1}!}{\qn{n}!\qn{b+c-n}!}q^{\pm bn}.
\end{equation}
 
We finish this preliminary part with a lemma that be useful in the next section, and whose proof could be found in \cite{ran-aag}.
\begin{Lem}\label{lem_aux1}
Let us consider three operators $A_+$, $A_-$, and $B$ such that $\cm{B}{A_\pm}=\pm A_\pm$, then
\begin{align}
\nonumber
B^\ell A_\pm &=A_\pm(B\pm I)^\ell, \\[10pt]
\label{lem2}\qn{\nu B+\eta}A_\pm^r&=A_\pm^r\qn{\nu B+\eta\pm\nu r}, \qquad \nu,\eta\in\mathbb{R},
\end{align}
for any $\ell,r\in\mathbb{Z^+}$. Moreover,
\begin{equation*}
\cm{B}{A_\pm^r}=\pm r A_\pm^r,
\end{equation*}
and
\begin{align}
\nonumber
&\cm{A_\pm}{A_\mp^r}=\pm A_\mp^{r-1}\qn{r}\qn{2B\mp(r-1)}
&\text{if}\quad \cm{A_\pm}{A_\mp}=\pm\qn{2B}, \\[4pt]
\label{lem5}
&\cm{A_\pm}{A_\mp^r}=\mp A_\mp^{r-1}\qn{r}\qn{2B\mp(r-1)}
&\text{if}\quad \cm{A_\pm}{A_\mp}=\mp\qn{2B}.
\end{align}
\end{Lem}

\section{The $su_q(1,1)$ algebra}
The present section is devoted to the development of basic facts about the $su_q(1,1)$ algebra in the spirit of the one carried out in \cite{ran-aag} for $su_q(2)$, but for a complete review the interested lector is urged to consult~\cite{smi-g}. Further on, we shall address the method of projection operator for $su_q(1,1)$ algebra by means of the ideas exhibited in \cite{smi1} for the $su_q(2)$ algebra (see also \cite{ran-aag}).

\subsection{Basic facts on the unitary representations of the $su_q(1,1)$ algebra}
The $su_q(1,1)$ algebra is generated by the operators $K_+$, $K_-$ and $K_0$, which fulfill the relations
$$
[K_0,K_\pm]=\pm K_\pm, \qquad [K_+,K_-]=-[2K_0],
$$
and the adjoint properties
$$
K_\pm^\dag=K_\mp, \qquad K_0^\dag=K_0.
$$
It is worth to pointing out that these relations came from the ones used in \cite{ran-mt} and different relations, but equivalent to the former, could be chosen; for example, see~\cite{gromov}.

The $su_q(1,1)$ is a non-compact algebra, so its non-trivial discrete irreducible and unitary representations are not finite-dimensional and besides there are two different possibilities: the so-called positive discrete series, whose spectrum is bounded below, and the negative one, whose spectrum is bounded above (vid.~Appendix \ref{ap-al} for the details). A basis of the irreducible and unitary representation of the positive discrete series will be denoted by $D^{\kappa+}$, and is given by 
$$
\ket{\kappa \mu}, \qquad \kappa=0,\frac{1}{2},1,\frac{3}{2},\ldots, \qquad \mu=\kappa+1,\kappa+2,\ldots,
$$
We should notice that the maximal weight vector does not exist and the minimal is $\ket{\kappa\,\kappa+1}$.

The action of the generators on the basis is given by
\begin{equation}\label{action}
\begin{split}
&K_0\ket{\kappa\mu}=\mu\ket{\kappa\mu}, \\[2pt]
&K_{-}\ket{\kappa\mu}=\sqrt{[\mu+\kappa]_q[\mu-\kappa-1]_q}\ket{\kappa\,\mu-1}, \\[2pt]
&K_+\ket{\kappa\mu}=\sqrt{[\mu-\kappa]_q[\mu+\kappa+1]_q}\ket{\kappa\,\mu+1}.
\end{split}
\end{equation}
Hence, the explicit form of the irreducible representation is determined by the equations
$$
\mel{\kappa\mu'}{K_0}{\kappa\mu}=\mu\delta_{\mu,\mu'}
\quad\text{and}\quad
\mel{\kappa\mu'}{K_\pm}{\kappa\mu}=\sqrt{[\mu\mp\kappa]_{q}[\mu\pm\kappa\pm1]_q}\delta_{\mu\pm1,\mu'}
$$
where $\delta_{\ell,k}$ denotes the usual Kronecker delta, which is one if $\ell=k$ and null otherwise.

Formulae for the iterated composition of each operator with itself could bee obtained by an induction procedure, which leads to
\begin{equation}\label{K^r}
\begin{split}
&K_{0}^{r}\ket{\kappa\mu}=\mu^r\ket{\kappa\mu} \\
&K_{-}^{r}\ket{\kappa\mu}=\sqrt{\frac{[\mu+\kappa]_q![\mu-\kappa-1]_q!}{[\mu+\kappa-r]_q![\mu-\kappa-1-r]_q!}}\ket{\kappa\,\mu-r},  \\
&K_{+}^{r}\ket{\kappa\mu}=\sqrt{\frac{[\mu+\kappa+r]_q![\mu-\kappa-1+r]_q!}{[\mu+\kappa]_q![\mu-\kappa-1]_q!}}\ket{\kappa\,\mu+r},
\end{split}\end{equation}
where $r\in\mathbb{Z}^{+}$. In particular, the matrix elements for powers of the generators are
\begin{align*}
\mel{\kappa\mu'}{K_{-}^{r}}{\kappa\mu}&=\sqrt{\frac{[\mu+\kappa]_q![\mu-\kappa-1]_q!}{[\mu+\kappa-r]_q![\mu-\kappa-1-r]_q!}}\delta_{\mu',\mu-r}, \\[5pt]
\mel{\kappa\mu'}{K_{+}^{r}}{\kappa\mu}&=\sqrt{\frac{[\mu+\kappa+r]_q![\mu-\kappa-1+r]_q!}{[\mu+\kappa]_q![\mu-\kappa-1]_q!}}\delta_{\mu',\mu+r}.
\end{align*}
In addition, an explicit formula for the basis vectors in terms of the minimal weight vector could be obtained; it is given by
$$
\ket{\kappa\mu}=\sqrt{\frac{[2\kappa+1]_q!}{[\mu+\kappa]_q![\mu-\kappa-1]_q!}}K_{+}^{\mu-\kappa-1}\ket{\kappa\,\kappa+1}.
$$

On its behalf, we define the Casimir operator of second order for the $su_q(1,1)$ algebra by
$$
C=-K_{+}K_{-}+[K_0]_q[K_0-1]_q=-K_{-}K_{+}+[K_0+1]_q[K_0]_q,
$$
which is self-adjoint, i.e., $C^{\dag}=C$. Its action on a basis vector is
$$
C\ket{\kappa\mu}=[\kappa+1]_q[\kappa]_q\ket{\kappa\mu},
$$
so all the family of vectors $\ket{\kappa\mu}$ are eigenvectors of the Casimir operator with the eigenvalue $[\kappa+1]_q[\kappa]_q$. Therefore, since it commutes with the generator $K_0$ and all $\ket{\kappa\mu}$ are also eigenvectors of $K_0$, both operators share a common orthonormal system of eigenvectors which is $\ket{\kappa\mu}$ itself, rather, 
$$
\braket{\kappa'\mu'}{\kappa\mu}=\delta_{\kappa,\kappa'}\delta_{\mu,\mu'}.
$$

\subsection{Projection operator for the $su_q(1,1)$ algebra}
Let $\ket{\kappa'\mu'}$ be any vector of the basis of an unitary irreducible representation of the positive discrete series kind $D^{\kappa'+}$ such that $\mu'=\kappa+1$, which implies $\kappa'\leq \kappa$.
We consider now the application $P_{\kappa+1,\kappa+1}^{\kappa+}$ from $D^{\kappa'+}$ to the vector space $\text{span}\{\ket{\kappa\,\kappa+1}\}$ defined by
\begin{equation}\label{def-po-su11}
P_{\kappa+1,\kappa+1}^{\kappa+}\ket{\kappa'\,\kappa+1}=\delta_{\kappa,\kappa'}\ket{\kappa\,\kappa+1}.
\end{equation}
Such application is called the projection operator for the $su_q(1,1)$ algebra, because from all possible vectors $\ket{\kappa'\,\kappa+1}$ belonging to the representation $D^{\kappa'+}$, it extracts the minimum weight vector $\ket{\kappa\,\kappa+1}$ of the representation $D^{\kappa+}$.

It is easy to check that the commutator of the projection operator and the generator $K_0$ is null when it is applied to a vector of the type $\ket{\kappa'\,\kappa+1}$, that is, $[P_{\kappa+1,\kappa+1}^{\kappa+},K_0]\ket{\kappa'\,\kappa+1}=0$. So, if we want to put the projection operator as an expansion in terms of the generators $K_{-}$ and $K_{+}$, they have to be raised to the same power. In this way, we set
$$
P_{\kappa+1,\kappa+1}^{\kappa+}=\sum_{r=0}^{\infty}c_{r}K_{+}^{r}K_{-}^{r},
$$
where $c_r\in\mathbb{R}$ are unknown at the moment. Notice that, in principle, the latter expansion is a series, however it is actually a sum whose last index value is some $r_0$ as it is written; for example, if we apply it to the vector $\ket{\kappa'\,\kappa+1}$, it is clear that all terms with $r>\kappa-\kappa'$
vanish.

Let us obtain an explicit expression for the previous coefficients $c_r$ in the way carried out in~\cite{ran-aag}. First, having in mind the observation done in the previous paragraph, it is clear the identity
$$
P_{\kappa+1,\kappa+1}^{\kappa+}\ket{\kappa\,\kappa+1}=c_0\ket{\kappa\,\kappa+1}
$$
which implies $c_0=1$ by comparison with \eqref{def-po-su11}. For the remainder coefficients we use   the fact that the composition of the projection operator and the generator $K_-$ applied to a vector of the type $\ket{\kappa'\,\kappa+1}$ is null, then, by identities \eqref{lem5} and \eqref{lem2} of Lemma \ref{lem_aux1}, we get
\begin{align*}
0=K_{-}P_{\kappa+1,\kappa+1}^{\kappa+}\ket{\kappa'\,\kappa+1}
&=\sum_{r=0}^{\infty}c_r K_{-}K_{+}^{r}K_{-}^{r}\ket{\kappa'\,\kappa+1} \\
&=\sum_{r=0}^{\infty}c_r(K_{+}^{r}K_{-}^{r+1}+K_{+}^{r-1}\qn{r}\qn{2K_{0}+r-1}K_{-}^{r})\ket{\kappa'\,\kappa+1} \\
&=\sum_{r=0}^{\infty}(c_r+c_{r+1}\qn{r+1}\qn{2K_0-r})K_{+}^{r}K_{-}^{r+1}\ket{\kappa'\,\kappa+1}.
\end{align*}
Therefore, by means of \eqref{K^r}, we have
$$
\sqrt{\qn{\kappa+\kappa'}!\qn{\kappa-\kappa'}!\qn{\kappa+\kappa'+1}!\qn{\kappa-\kappa'-1}!}\sum_{r=0}^{\infty}
\frac{ ( c_r+c_{r+1}\qn{r+1}\qn{2\kappa-r}  ) }{\qn{\kappa+\kappa'-r}!\qn{\kappa-\kappa'-1-r}!}\ket{\kappa'\kappa}=0.
$$
Last equation naturally implies
$$
\sum_{r=0}^{\infty}\frac{c_r+c_{r+1}[r+1]_q[2\kappa-r]_q}{[\ell-1-r]_q![2\kappa-\ell-r]_q!}\ket{\kappa-\ell\,\kappa}=0
$$
and, by means of it, we deduce through a recurrence process that the coefficients are
$$
c_r=\frac{(-1)^{r}[2\kappa-r]_q!}{[r]_q![2\kappa]_q!} \text{  \quad if $r\leq \kappa$ and $0$ otherwise.}
$$
So, finally, the desired series expansion for the projection operator is given by
$$
P_{\kappa+1,\kappa+1}^{\kappa+}=\sum_{r=0}^{\infty}\frac{(-1)^{r}[2\kappa-r]_q!}{[r]_q![2\kappa]_q!}K_{+}^{r}K_{-}^{r}.
$$
It must be taken into account that when applying the above series to the vectors $\ket{\kappa',\kappa+1}$, only a finite number of terms are non-vanishing, so there are no convergence issues.  This is, in fact, the situation that one finds when one uses the projection operators for calculating the Clebsch-Gordan coefficients as it will be shown latter on.

A generalized projection operator could be defined in the same way as before. Let us consider any vector $\ket{\kappa'\mu'}$ of  the unitary irreducible representation of the positive discrete series $D^{\kappa'+}$ such that $\mu'\geq \kappa+1$, which also implies $\kappa'\leq \kappa$. We define the generalized projection operator $P_{\mu,\mu'}^{\kappa+}$ from $D^{\kappa+}$ to the vector space $\text{span}\{\ket{\kappa\mu}\}$ by
\begin{equation}\label{su11_proj-gen}
P_{\mu,\mu'}^{\kappa+}=\sqrt{\frac{[2\kappa+1]_q!}{[\mu+\kappa]_q![\mu-\kappa-1]_q!}}K_{+}^{\mu-\kappa-1}P_{\kappa+1,\kappa+1}^{\kappa+}K_{-}^{\mu'-\kappa-1}\sqrt{\frac{[2\kappa+1]_q!}{[\mu'+\kappa]_q![\mu'-\kappa-1]_q!}},
\end{equation}
which  of course coincides with the projection operator $P_{\kappa+1\,\kappa+1}^{\kappa+}$ defined previously when $\mu'=\kappa+1$. However, in contrast with the  latter, the generalized one is not self-adjoint but it satisfies the properties $(P_{\mu,\mu'}^{\kappa+})^\dag=P_{\mu',\mu}^{\kappa+}$ and
\begin{equation}\label{poxpo}
P_{\mu,\mu'}^{\kappa+}P_{\mu',\mu''}^{\kappa+}=P_{\mu,\mu''}^{\kappa+}.
\end{equation}
Furthermore, by a direct calculation it is easy to prove that $P_{\mu,\mu'}^{\kappa+}\ket{\kappa'\mu'}=\delta_{\kappa,\kappa'}\ket{\kappa\mu}$.

An important fact related to this operator is that the action on a linear combination of a basis vectors $\ket{\kappa'\mu'}$, where $\mu'$ has a fixed value, is proportional to $\ket{\kappa\mu}$. Indeed, if $\ket{\cdot\,\mu'}=\sum_{\kappa'}a_{\kappa'\mu'}\ket{\kappa'\mu'}$ with $a_{\kappa'\mu'}\in\mathbb{R}$ not all simultaneously zero, then
$$
P_{\mu,\mu'}^{\kappa+}\ket{\cdot\ \mu'}=\sum_{\kappa'}a_{\kappa'\mu'}P_{\mu,\mu'}^{\kappa+}\ket{\kappa'\mu'}=a_{\kappa\mu'}\ket{\kappa\mu}.
$$

\section{The Clebsch-Gordan coefficients}\label{sec:cgcReps}
Let us consider the direct product $D^{\kappa_1}\otimes D^{\kappa_2}$ of two representations $D^{\kappa_1}$ and $D^{\kappa_2}$ of the $su_q(1,1)$ algebra. In general, the direct product of two representations can be expressed as a direct sum of the irreducible representations, i.e., 
$$
D^{\kappa_1}\otimes D^{\kappa_2}=\bigoplus_{\kappa} D^{\kappa},
$$
where the sums runs in a certain set of discrete values related with $\kappa_1$ and $\kappa_2$ as we shall show later on. In the present work we do not consider the case of a continuous spectrum.

Let  $\ket{\kappa_1\mu_1}$ and $\ket{\kappa_2\mu_2}$ the orthogonal basis vectors of $D^{\kappa_1}$ and $D^{\kappa_2}$, respectively, and let $\ket{\kappa_1,\kappa_2;\kappa\mu}$ be basis vectors of $D^{\kappa}$. Then, a typical situation occurs when the vectors $\ket{\kappa_1,\kappa_2;\kappa\mu}$ are expanded in terms of the basis $\ket{\kappa_1\mu_1}\ket{\kappa_2\mu_2}$, that is,
\begin{equation}\label{jm-jimi}
\ket{\kappa_1,\kappa_2;\kappa\mu}=\sum_{\substack{\mu_1,\,\mu_2 \\ \mu=\mu_1+\mu_2}}C_{\kappa\mu}^{\kappa_1\mu_1,\kappa_2\mu_2}\ket{\kappa_1\mu_1}\ket{\kappa_2\mu_2}
=\sum_{\substack{\mu_1,\, \mu_2 \\ \mu=\mu_1+\mu_2}}\braket{\kappa_1\mu_1,\kappa_2\mu_2}{\kappa\mu}_{q}\ket{\kappa_1\mu_1}\ket{\kappa_2\mu_2}.
\end{equation}
The coefficients $C_{\kappa\mu}^{\kappa_1\mu_1,\kappa_2\mu_2}$ in the previous expansion are called the Clebsh-Gordan coefficients and are usually denoted by $\braket{\kappa_1\mu_1,\kappa_2\mu_2}{\kappa\mu}_{q}$. Our main aim in the following sections is to compute them in a simple way in two different situations: first, if both $D^{\kappa_1}$ and $D^{\kappa_2}$ are irreducible and unitary representations of discrete series, and second, if $D^{\kappa_1}$ is irreducible and unitary but $D^{\kappa_2}$ is non-irreducible (finite-dimensional). To this end, we notice that an analogous previous argument could be done for the basis $\ket{\kappa_1\mu_1}\ket{\kappa_2\mu_2}$ expanding it in terms of $\ket{\kappa_1,\kappa_2;\kappa\mu}$, so
\begin{equation*}
\ket{\kappa_1\mu_1}\ket{\kappa_2\mu_2}=\sum_{\substack{\bar{\kappa},\,\bar{\mu} \\ \bar{\mu}=\mu_1+\mu_2 \\ \bar{\kappa}=\bar{\kappa}(\kappa_1,\kappa_2)}}\bar{C}_{\bar{\kappa},\bar{\mu}}^{\kappa_1\bar{\mu}_1,\kappa_2\bar{\mu}_2}
\ket{\kappa_1,\kappa_2;\bar{\kappa}\bar{\mu}}.
\end{equation*}
Since the basis vectors $\ket{\kappa_1\mu_1}$ and $\ket{\kappa_2\mu_2}$ can be always assume to be orthonormal then, from \eqref{jm-jimi},  the identity
$$
\braket{\kappa_1\mu_1,\kappa_2\mu_2}{\kappa\mu}_{q}=\bra{\kappa_1\mu_1}\bra{\kappa_2\mu_2}\cdot \ket{\kappa_1,\kappa_2;\kappa\mu}
$$
holds. But we  already know that applying the  generalized projection operator $P^{\kappa+}_{\mu,\bar{\mu}}$ to a linear combination (in $\bar{\kappa}$) of vectors $\ket{\kappa_1,\kappa_2;\bar{\kappa}\bar{\mu}}$ one obtains a vector proportional to $\ket{\kappa_1,\kappa_2;\kappa\mu}$, thus
$$
P^{\kappa+}_{\mu,\bar{\mu}}\ket{\kappa_1\bar{\mu}_1}\ket{\kappa_2\bar{\mu}_2}= a_{\kappa\bar{\mu}}\ket{\kappa_1,\kappa_2;\kappa\mu}
\iff
\ket{\kappa_1,\kappa_2;\kappa\mu}=\frac{P^{\kappa+}_{\mu,\bar{\mu}}\ket{\kappa_1\bar{\mu}_1}\ket{\kappa_2\bar{\mu}_2}}{\|P^{\kappa+}_{\mu,\bar{\mu}}\ket{\kappa_1\bar{\mu}_1}\ket{\kappa_2\bar{\mu}_2}\|},
$$
where, since, in our case  $(P^{\kappa+}_{\mu,\bar{\mu}})^\dag=\proj{\kappa+}{\bar{\mu}\,\mu}$ and using \eqref{poxpo} we find
\begin{equation}\label{norm_dot}
\begin{split}
\|P^{\kappa+}_{\mu,\bar{\mu}}\ket{\kappa_1\bar{\mu}_1}\ket{\kappa_2\bar{\mu}_2}\|^2
&=\bra{\kappa_1\bar{\mu}_1}\bra{\kappa_2\bar{\mu}_2}(P^{\kappa+}_{\mu,\bar{\mu}})^{\dag}P^{\kappa+}_{\mu,\bar{\mu}}\ket{\kappa_1\bar{\mu}_1}\ket{\kappa_2\bar{\mu}_2} \\
&= \bra{\kappa_1\bar{\mu}_1}\bra{\kappa_2\bar{\mu}_2}P^{\kappa+}_{\bar{\mu},\bar{\mu}}\ket{\kappa_1\bar{\mu}_1}\ket{\kappa_2\bar{\mu}_2}
\end{split}
\end{equation}
if we want $\ket{\kappa_1,\kappa_2;\kappa\mu}$ to be normalized to one.

From the above reasoning it follows that the Clebsch-Gordan coefficients can be written in terms of the projection operator by the expression
\begin{equation}\label{eq_GenCG}
\braket{\kappa_1\mu_1,\kappa_2\mu_2}{\kappa\mu}_{q}=
\frac{\bra{ \kappa_1\mu_1}\bra{\kappa_2\mu_2}P^{\kappa+}_{\mu,\bar{\mu}}\ket{\kappa_1\bar{\mu}_1}\ket{\kappa_2\bar{\mu}_2}}
{\|P^{\kappa+}_{\mu,\bar{\mu}}\ket{\kappa_1\bar{\mu}_1}\ket{\kappa_2\bar{\mu}_2}\|}.
\end{equation}

It is a matter of fact that both $\bar{\mu}_1$ and $\bar{\mu}_2$ are free parameters, so we can choose the most appropriate values in each case to make easy the calculation of the Clebsch-Gordan coefficients.  This is the key point and where the full capability of the method reveals.

\subsection{Clebsch-Gordan coefficients for the product of two irreducible and unitary representations}\label{subsec-CGCIR}
Let us consider two different irreducible and unitary representations for the positive discrete series of order $\kappa_1$ and $\kappa_2$ of the $su_q(1,1)$ algebra, i.e.,
$$
D^{\kappa_i+}, \qquad \ket{\kappa_i\mu_i}, \quad \kappa=0,\frac{1}{2},1,\frac{3}{2},.\ldots, \quad \mu_i=\kappa_i+1,\kappa_i+2,\ldots, 
$$
for $i=1,2$. 

The direct product representation $D^{\kappa_1+}\otimes D^{\kappa_2+}$ has a basis $\ket{\kappa_1\mu_1}\ket{\kappa_2\mu_2}$ 
and it is generated by the operators
$$
K_{0}(12)=K_{0}\otimes1+1\otimes K_{0}
\quad\text{and}\quad
K_{\pm}(12)=K_{\pm}\otimes q^{K_0}+q^{-K_0}\otimes K_{\pm},
$$
but from the sake of clarity we use the notation
$$
K_0(12)=K_0(1)+K_0(2) \quad\text{and}\quad K_\pm(12)=K_\pm(1)q^{K_0(2)}+q^{-K_0(1)}K_\pm(2).
$$
where the number in brackets indicates the action of the operator on the basis of the product representation.  The appearance of the factors $q^{K_0}$ and $q^{-K_0}$ in the generators of the direct product arise from the lack of linearity of the quantum number and they satisfy the properties
$$
q^{aK_0}K_0=K_0q^{aK_0} \quad\text{and}\quad q^{aK_0}K_\pm=K_\pm q^{a(K_0\pm1)}.
$$
Additionally, same considerations are obtained if the operators
$$
\tilde{K}_{0}(12)=K_{0}\otimes 1+1\otimes K_{0}
\quad\text{and}\quad
\tilde{K}_{\pm}(12)=K_{\pm}\otimes q^{-K_0}+q^{K_0}\otimes K_{\pm}
$$
were defined with the minor change $q^{-1}$ instead of~$q$.

The generators satisfy both the adjointness properties $(K_0(12))^\dag = K_0(12)$ and $(K_\pm(12))^\dag = K_\mp(12)$, as well as the commutation properties
$$
\cm{K_0(12)}{K_\pm(12)}=\pm K_\pm(12)
\quad\text{and}\quad
\cm{K_+(12)}{K_-(12)}=-\qn{2K_0(12)}.
$$
Note that in this situation the statements of the Lemma \ref{lem_aux1} can be used with the operators $K_{-}(12)$, $K_{+}(12)$ and $K_0(12)$.

Moreover, from the above relations it is easy to check that 
$$
K_0^r(12)\ket{\kappa_1\mu_1}\ket{\kappa_2\mu_2}=(\mu_1+\mu_2)^r\ket{\kappa_1\mu_1}\ket{\kappa_2\mu_2},
$$
as well as
\begin{equation}\label{eq11_BinExp}
K_\pm^r(12)=\sum_{\ell=0}^{r}\frac{[r]_q!}{[\ell]_q![r-\ell]_q!}K_\pm^\ell(1)K_\pm^{r-\ell}(2)
q^{\ell K_0(2)-(r-\ell)K_0(1)}
\end{equation}
Also we can built the Casimir operator for the resulting representation that, in this case, reads
$$
C(12)=-K_{+}(12)K_{-}(12)+\qn{K_0(12)}\qn{K_0-1}
=-K_{-}(12)K_{+}(12)+\qn{K_0(12)+1}\qn{K_0(12)}.
$$
At this point we are in position to compute the Clebsch-Gordan coefficients. To this end, we use the general formula \eqref{eq_GenCG} where, as it was noted there, the choice of the parameters $\bar{\mu}_1$ and $\bar{\mu}_2$ is free, so we take $\bar{\mu}_1=\kappa_1+1$ and $\bar{\mu}_2=\kappa-\kappa_1$, which implies $\bar{\mu}=\bar{\mu}_1+\bar{\mu}_2=\kappa+1$. The reasons for this decision will be clear soon. Therefore
\begin{equation}\label{ccg-su11}
\braket{\kappa_1\mu_1,\kappa_2\mu_2}{\kappa\mu}=\frac{
\bra{\kappa_1\mu_1}\bra{\kappa_2\mu_2}P^{\kappa+}_{\mu,\kappa+1}\ket{\kappa_1\,\kappa_1+1}\ket{\kappa_2\,\kappa-\kappa_1}}
{\|P^{\kappa+}_{\mu,\kappa+1}\ket{\kappa_1\,\kappa_1+1}\ket{\kappa_2\,\kappa-\kappa_1}\|}. 
\end{equation}
where $\kappa\geq\kappa_1+\kappa_2+1$, since $\kappa_2+1\leq\bar{\mu}_2=\kappa-\kappa_1$.

The main tool to do the calculation of the numerator in the expression \eqref{ccg-su11} is the generalized projection operator showed in \eqref{su11_proj-gen}. Here, the advantage of our precious selection in the parameters appears clearly, because its expression is the quite simply
\begin{align*}
P^{\kappa+}_{\mu,\kappa+1}
&=\sqrt{\frac{[2\kappa+1]_q!}{[\mu+\kappa]_q![\mu-\kappa-1]_q!}}K_{+}^{\mu-\kappa-1}(12)P^{\kappa+}_{\kappa+1,\kappa+1} \\
&=\sqrt{\frac{[2\kappa+1]_q!}{[\mu+\kappa]_q![\mu-\kappa-1]_q!}}
\sum_{r=0}^{\infty}\frac{(-1)^r[2\kappa-r]_q!}{[r]_q![2\kappa]_q!}K_{+}^{r+\mu-\kappa-1}(12)K_{-}^{r}(12),
\end{align*}
Therefore, the numerator of \eqref{su11_proj-gen} is given by
$$
\bra{\kappa_1\mu_1}\bra{\kappa_2\mu_2}
\sqrt{\frac{[2\kappa+1]_q!}{[\mu+\kappa]_q![\mu-\kappa-1]_q!}}
\sum_{r=0}^{\infty}\frac{(-1)^r[2\kappa-r]_q!}{[r]_q![2\kappa]_q!}K_{+}^{\mu-\kappa-1+r}(12)K_{-}^{r}(12)
\ket{\kappa_1\,\kappa_1+1}\ket{\kappa_2\,\kappa-\kappa_1}
$$
By means of the binomial expansion \eqref{eq11_BinExp} we obtain
\begin{multline*}
K_{-}^{r}(12)\ket{\kappa_1\kappa_1+1}\ket{\kappa_2\kappa-\kappa_1}=\sum_{\ell=0}^{r}\frac{[r]_q!}{[\ell]_q![r-\ell]_q!}
q^{\ell(\kappa-\kappa_1)-(r-\ell)(\kappa_1+1)}
\sqrt{\frac{[2\kappa_1+1]_q!}{[2\kappa_1+1-\ell]_q![-\ell]_q!}}
\\
\times\sqrt{\frac{[\kappa-\kappa_1+\kappa_2]_q![\kappa-\kappa_1-\kappa_2-1]_q!}{[\kappa-\kappa_1+\kappa_2-r+\ell]_q![\kappa-\kappa_1-\kappa_2-1-r+\ell]_q!}}
\ket{\kappa_1\,\kappa_1+1-\ell}\ket{\kappa_2\, \kappa-\kappa_1-r+\ell}
\end{multline*}
but, since we have chosen $\bar{\mu}_1$ as the minimum value $\kappa_1+1$, the only possible addend in the sum is the one corresponding with $\ell=0$, which implies
\begin{multline*}
K_{-}^{r}(12)\ket{\kappa_1\kappa_1+1}\ket{\kappa_2\kappa-\kappa_1}=
\sqrt{\frac{[\kappa-\kappa_1+\kappa_2]_q![\kappa-\kappa_1-\kappa_2-1]_q!}{[\kappa-\kappa_1+\kappa_2-r]_q![\kappa-\kappa_1-\kappa_2-1-r]_q!}}
\\
\times q^{-r(\kappa_1+1)}\ket{\kappa_1\,\kappa_1+1}\ket{\kappa_2\, \kappa-\kappa_1-r}.
\end{multline*}
Then, having in mind the inequality $\kappa-\kappa_1-\kappa_2-1<\kappa-\kappa_1+\kappa_2$, we get for the numerator the expression
\begin{multline*}
\sqrt{\frac{[2\kappa+1]_q!}{[\mu+\kappa]_q![\mu-\kappa-1]_q!}}\sum_{r=0}^{\kappa-\kappa_1-\kappa_2-1}\frac{(-1)^r[2\kappa-r]_q!}{[r]_q![2\kappa]_q!}\sqrt{\frac{[\kappa-\kappa_1+\kappa_2]_q![\kappa-\kappa_1-\kappa_2-1]_q!}{[\kappa-\kappa_1+\kappa_2-r]_q![\kappa-\kappa_1-\kappa_2-1-r]_q!}}
\\
\times q^{-r(\kappa_1+1)}\bra{\kappa_1\mu_1}\bra{\kappa_2\mu_2}K_{+}^{\mu-\kappa-1+r}(12)\ket{\kappa_1\,\kappa_1+1}\ket{\kappa_2\, \kappa-\kappa_1-r}.
\end{multline*}
Using again the binomial expansion \eqref{eq11_BinExp} we obtain
\begin{multline*}
K_{+}^{\mu-\kappa-1+r}(12)\ket{\kappa_1\,\kappa_1+1}\ket{\kappa_2\,\kappa-\kappa_1-r}\!\!
=\!\!\sum_{\ell=0}^{\mu-\kappa-1+r}\!\!\frac{[\mu\!-\!\kappa\!-\!1+r]_q!}{[\ell]_q![\mu-\kappa-1+r-\ell]_q!}
\sqrt{\frac{[2\kappa_1\!+\!1\!+\!\ell]_q![\ell]_q!}{[2\kappa_1+1]_q!}} \\
\times\sqrt{\frac{[\mu-\kappa_1+\kappa_2-1-\ell]_q![\mu-\kappa_1-\kappa_2-2-\ell]_q!}{[\kappa-\kappa_1+\kappa_2-r]_q![\kappa-\kappa_1-\kappa_2-1-r]_q!}} 
q^{(\kappa-\kappa_1-r)\ell-(\mu-\kappa-1+r-\ell)(\kappa_1+1)} \\
\times\ket{\kappa_1\,\kappa_1+1+\ell}\ket{\kappa_2\,\mu-\kappa_1-1-\ell},
\end{multline*}
that leads to
\begin{multline*}
\sqrt{\frac{[2\kappa+1]_q[\kappa-\kappa_1+\kappa_2]_q![\kappa-\kappa_1-\kappa_2-1]_q!}{[2\kappa]_q![\mu+\kappa]_q![\mu-\kappa-1]_q![2\kappa_1+1]_q!}}q^{(\kappa+1-\mu)(\kappa_1+1)} \\
\times\sum_{r=0}^{\kappa-\kappa_1-\kappa_2-1}\frac{(-1)^r[2\kappa-r]_q![\mu-\kappa-1+r]_q!}{[r]_q![\kappa-\kappa_1+\kappa_2-r]_q![\kappa-\kappa_1-\kappa_2-1-r]_q!}q^{-2(\kappa_1+1)r} \\
\times\sum_{\ell=0}^{\mu-\kappa-1+r}\frac{\sqrt{[2\kappa_1+1+\ell]_q![\mu-\kappa_1+\kappa_2-1-\ell]_q![\mu-\kappa_1-\kappa_2-2-\ell]_q![\ell]_q!}}{[\ell]_q![\mu-\kappa-1+r-\ell]_q!}q^{(\kappa+1-r)\ell} \\
\times\braket{\kappa_1\mu_1}{\kappa_1\,\kappa_1+1+\ell}\braket{\kappa_2\mu_2}{\kappa_2\,\mu-\kappa_1-1-\ell}.
\end{multline*}
The orthogonality of the basis implies that $\ell=\mu_1-\kappa_1-1$, which, in turn, means that all the summands in the last formula vanish if $\kappa-\kappa_1-\mu_2\leq r$. Consequently, for the numerator of \eqref{ccg-su11}, we find that
\begin{multline*}
\bra{\kappa_1\mu_1}\bra{\kappa_2\mu_2}P^{\kappa+}_{\mu,\kappa+1}\ket{\kappa_1\,\kappa_1+1}\ket{\kappa_2\,\kappa-\kappa_1}=q^{\mu_1(\kappa+1)-\mu(\kappa_1+1)} \\
\times\sqrt{\frac{[2\kappa+1]_q[\kappa-\kappa_1+\kappa_2]_q![\kappa-\kappa_1-\kappa_2-1]_q![\mu_1+\kappa_1]_q![\mu_2+\kappa_2]_q![\mu_2-\kappa_2-1]_q!}{[2\kappa]_q![\mu+\kappa]_q![\mu-\kappa-1]_q![2\kappa_1+1]_q![\mu_1-\kappa_1-1]_q!}} \\
\times\sum_{r=\max\{0,\kappa-\kappa_1-\mu_2\}}^{\kappa-\kappa_1-\kappa_2-1}\frac{(-1)^r[2\kappa-r]_q![\mu-\kappa-1+r]_q!}{[r]_q![\kappa-\kappa_1+\kappa_2-r]_q![\kappa-\kappa_1-\kappa_2-1-r]_q![\mu_2+\kappa_1-\kappa+r]_q!}q^{-(\mu_1+\kappa_1+1)r}.
\end{multline*}
Notice  that, if $\kappa-\kappa_1-\mu_2>0$,  and taking into account the identity
$$
\frac{1}{[\mu_2+\kappa_1-\kappa+r]_q!}=(-1)^{\kappa-\kappa_1-\mu_2}\frac{(-r|q)_{\kappa-\kappa_1-\mu_2}}{[r]_q!},
$$
the lower index of the sum in the above expression can be taken equal zero because all the extra terms will vanish.

On its behalf, the denominator of the Clebsch-Gordan coefficient in the same expression \eqref{ccg-su11} could be computed having in mind the equation \eqref{norm_dot}, which in the present setting is
$$
\|P_{\mu,\kappa+1}^{\kappa+}\ket{\kappa_1\,\kappa_1+1}\ket{\kappa_2\,\kappa-\kappa_1}\|^{2}
=\bra{\kappa_1\,\kappa_1+1}\bra{\kappa_2\,\kappa-\kappa_1}P_{\mu,\kappa+1}^{\kappa+}\ket{\kappa_1\,\kappa_1+1}\ket{\kappa_2\,\kappa-\kappa_1},
$$
and the value of the numerator with the choice of parameters $\mu_1=\kappa_1+1$ and $\mu_2=\kappa-\kappa_1$. Then
\begin{multline*}
\|P_{\mu,\kappa+1}^{\kappa+}\ket{\kappa_1\,\kappa_1+1}\ket{\kappa_2\,\kappa-\kappa_1}\|^{2}
=\frac{[\kappa-\kappa_1+\kappa_2]_q![\kappa-\kappa_1-\kappa_2-1]_q!}{[2\kappa]_q!} \\
\times\sum_{r=0}^{\kappa-\kappa_1-\kappa_2-1}\frac{(-1)^{r}[2\kappa-r]_q!}{[r]_q![\kappa-\kappa_1+\kappa_2-r]_q![\kappa-\kappa_1-\kappa_2-1-r]_q!}q^{-2(\kappa_1+1)r}.
\end{multline*}
But the sum could be putting into a series which can be summed by means of \eqref{suma2equiv} with the parameters $n=\kappa-\kappa_1-\kappa_2-1$, $b=\kappa-\kappa_1+\kappa_2$ and $c=2\kappa$  (note $b<c$), so
\begin{multline*}
\sum_{r=0}^{\kappa-\kappa_1-\kappa_2-1}\frac{(-1)^{r}[2\kappa-r]_q!}{[r]_q![\kappa-\kappa_1+\kappa_2-r]_q![\kappa-\kappa_1-\kappa_2-1-r]_q!}q^{-2(\kappa_1+1)r} \\
=\frac{[\kappa+\kappa_1-\kappa_2]_q![\kappa+\kappa_1+\kappa_2+1]_q!}{[\kappa-\kappa_1-\kappa_2-1]_q![\kappa-\kappa_1+\kappa_2]_q![2\kappa_1+1]_q!}q^{
(\kappa-\kappa_1+\kappa_2)(\kappa-\kappa_1-\kappa_2-1)}.
\end{multline*}
Hence, the expression for the denominator is
$$
\|P_{\mu,\kappa+1}^{\kappa+}\ket{\kappa_1\,\kappa_1+1}\ket{\kappa_2\,\kappa-\kappa_1}\|
=\sqrt{\frac{[\kappa+\kappa_1-\kappa_2]_q![\kappa+\kappa_1+\kappa_2+1]_q!}{[2\kappa]_q![2\kappa_1+1]_q!}}q^{(\kappa-\kappa_1+\kappa_2)(\kappa-\kappa_1-\kappa_2-1)/2}.
$$

Therefore, the Clebsch-Gordan coefficients of the product of two irreducible and unitary representations of $su_q(1,1)$ are given by
\begin{multline}\label{sum-alb}
\braket{\kappa_1\mu_1,\kappa_2\mu_2}{\kappa\mu}_q=
\sqrt{\frac{[2\kappa+1]_q[\kappa-\kappa_1+\kappa_2]_q![\kappa-\kappa_1-\kappa_2-1]_q!}{[\kappa+\kappa_1-\kappa_2]_q![\kappa+\kappa_1+\kappa_2+1]_q!}}q^{-(\kappa-\kappa_1+\kappa_2)(\kappa-\kappa_1-\kappa_2-1)/2} \\
\times\sqrt{\frac{[\mu_1+\kappa_1]_q![\mu_2+\kappa_2]_q![\mu_2-\kappa_2-1]_q!}{[\mu+\kappa]_q![\mu-\kappa-1]_q![\mu_1-\kappa_1-1]_q!}}q^{\mu_1(\kappa+1)-\mu(\kappa_1+1)} \\
\times\sum_{r=0}^{\kappa-\kappa_1-\kappa_2-1}\frac{(-1)^r[2\kappa-r]_q![\mu-\kappa-1+r]_q!}{[r]_q![\kappa-\kappa_1+\kappa_2-r]_q![\kappa-\kappa_1-\kappa_2-1-r]_q![\mu_2+\kappa_1-\kappa+r]_q!}q^{-(\mu_1+\kappa_1+1)r}.
\end{multline}

It is worth to indicate that a different expression could be obtained by reversing the order of summation in the latter one by means of the change of variable $r=\kappa-\kappa_1-\kappa_2-1$. Accordingly, the formula
\begin{multline}\label{sum-smi}
\braket{\kappa_1\mu_1,\kappa_2\mu_2}{\kappa\mu}_q=
(-1)^{\kappa-\kappa_1-\kappa_2-1}\sqrt{\frac{[2\kappa+1]_q[\kappa-\kappa_1+\kappa_2]_q![\kappa-\kappa_1-\kappa_2-1]_q!}{[\kappa+\kappa_1-\kappa_2]_q![\kappa+\kappa_1+\kappa_2+1]_q!}} \\
\times q^{-(\kappa+\kappa_1+\kappa_2+2)(\kappa-\kappa_1-\kappa_2-1)/2}\sqrt{\frac{[\mu_1+\kappa_1]_q![\mu_2+\kappa_2]_q![\mu_2-\kappa_2-1]_q!}{[\mu+\kappa]_q![\mu-\kappa-1]_q![\mu_1-\kappa_1-1]_q!}}q^{\mu_1(\kappa_2+1)-\mu_2(\kappa_1+1)} \\
\times\sum_{r=0}^{\min\{\kappa-\kappa_1-\kappa_2+1,\mu_2-\kappa_2-1\}}\frac{(-1)^r[\kappa+\kappa_1+\kappa_2+1+r]_q![\mu-\kappa_1-\kappa_2-2-r]_q!}{[r]_q![\kappa-\kappa_1-\kappa_2-1-r]_q![2\kappa_2+1+r]_q![\mu_2-\kappa_2-1-r]_q!}q^{(\mu_1+\kappa_1+1)r}
\end{multline}
is obtained, where we have recovered the original variable $r$. We should advise here that it matches the one appearing in \cite[Eq.~(5.576), p.~268]{gromov}.
In fact, the sum in the last expression can be taken over all non-negative integers 
because all the extra terms will vanish.

\subsubsection{Representations as a terminating $q$-hypergeometric series} Both sums in the two different expressions obtained for the Clebsch-Gordan coefficients could be extended to a series, since the add-on terms are all of them null. This fact allows us to transform both formulae to another different ones involving a symmetric terminating $q$-hypergeometric function $_{3}F_2$, which was defined by \eqref{q-hip-def}. To this end, we have to fix a nonnegative integer $n$ in each case. For example, $n=\kappa-\kappa_1-\kappa_2-1$ is the natural choice in the expression \eqref{sum-alb}, so
\begin{multline}\label{3f2-al}
\braket{\kappa_1\mu_1,\kappa_2\mu_2}{\kappa\mu}_q=
\sqrt{\frac{[2\kappa+1]_q![2\kappa]_q!q^{-(\kappa-\kappa_1+\kappa_2)(\kappa-\kappa_1-\kappa_2-1)} }{[\kappa+\kappa_1-\kappa_2]_q![\kappa-\kappa_1+\kappa_2]_![\kappa-\kappa_1-\kappa_2-1]_q![\kappa+\kappa_1+\kappa_2+1]_q!}} \\
\times\frac{1}{[\mu_2+\kappa_1-\kappa]_q!}\sqrt{\frac{[\mu-\kappa-1]_q![\mu_1+\kappa_1]_q![\mu_2+\kappa_2]_q![\mu_2-\kappa_2-1]_q!}{[\mu+\kappa]_q![\mu_1-\kappa_1-1]_q!}}
q^{\mu_1(\kappa+1)-\mu(\kappa_1+1)} \\
\times
 {_3}F_2 \bigg(\!\begin{array}{c} -\kappa+\kappa_1+\kappa_2+1 \, , \, \kappa_1-\kappa_2-\kappa \, ,\, \mu-\kappa \\
-2\kappa \, , \, \mu_2+\kappa_1-\kappa+1
 \end{array}\!\bigg\vert\,  q\, , \, q^{-(\mu_1+\kappa_1+1)}\!\bigg).
\end{multline}
To proceed analogously with formula \eqref{sum-smi}, we may choose  $n=\kappa-\kappa_1-\kappa_2+1$ or $n=\mu_2-\kappa_2-1$  whichever is smaller. Recall that all terms corresponding to summation indices greater than the minimum of these two values vanish. Then,
\begin{multline}\label{3f2-smi}
\braket{\kappa_1\mu_1,\kappa_2\mu_2}{\kappa\mu}_q=
\frac{(-1)^{\kappa-\kappa_1-\kappa_2-1}}{[2\kappa_2+1]_q!}
\sqrt{\frac{[2\kappa+1]_q[\kappa-\kappa_1+\kappa_2]_q![\kappa+\kappa_1+\kappa_2+1]_q!}{[\kappa+\kappa_1-\kappa_2]_q![\kappa-\kappa_1-\kappa_2-1]_!}} \\
\times q^{-(\kappa+\kappa_1+\kappa_2+2)(\kappa-\kappa_1-\kappa_2-1)/2}[\mu-\kappa_1-\kappa_2-2]_q!\sqrt{\frac{[\mu_1+\kappa_1]_q![\mu_2+\kappa_2]_q!}{[\mu+\kappa]_q![\mu-\kappa-1]_q![\mu_1-\kappa_1-1]_q![\mu_2-\kappa_2-1]}} \\
\times q^{\mu_1(\kappa+1)-\mu(\kappa_1+1)} 
 {_3}F_2 \bigg(\!\begin{array}{c} -\kappa+\kappa_1+\kappa_2+1 \, , \, \kappa+\kappa_1+\kappa_2+2 \, ,\, \kappa_2+1-\mu_2 \\
2\kappa_2+2 \, , \, \kappa_1+\kappa_2+2-\mu
 \end{array}\!\bigg\vert\,  q\, , \, q^{\mu_1+\kappa_1+1}\!\bigg).
\end{multline}

This approach to show the Clebsch-Gordan coefficients lets us to take advantage of the properties of the symmetric terminating $q$-hypergeometric function and relations between them.  For instance, if in formula \eqref{3f2-al} we take 
$$
n=\kappa-\kappa_1-\kappa_2-1,\ a=\mu-\kappa,\ b=\kappa_1-\kappa_2-\kappa,\ d=-2\kappa\ \text{and}\ e=\mu_2+\kappa_1-\kappa+1
$$ 
and use the transformation \eqref{142q}, we obtain the expression
\begin{multline*}
\braket{\kappa_1\mu_1,\kappa_2\mu_2}{\kappa\mu}_q=
(-1)^{\kappa-\kappa_1-\kappa_2-1}\sqrt{\frac{[2\kappa+1]_q![2\kappa]_q!q^{-(\kappa-\kappa_1+\kappa_2)(\kappa-\kappa_1-\kappa_2-1)} }{[\kappa+\kappa_1-\kappa_2]_q![\kappa-\kappa_1+\kappa_2]_![\kappa-\kappa_1-\kappa_2-1]_q!}} \\
\times\frac{1}{[\mu_1+\kappa_2-\kappa]_q!}\sqrt{\frac{[\mu-\kappa-1]_q![\mu_1+\kappa_1]_q![\mu_2+\kappa_2]_q![\mu_1-\kappa_1-1]_q!}{[\kappa+\kappa_1+\kappa_2+1]_q![\mu+\kappa]_q![\mu_2-\kappa_2-1]_q!}}
q^{\mu(\kappa_2+1)-\mu_2(\kappa+1)} \\
\times
 {_3}F_2 \bigg(\!\begin{array}{c} -\kappa+\kappa_1+\kappa_2+1 \, , \, \kappa_1-\kappa_2-\kappa \, ,\, \mu-\kappa \\
-2\kappa \, , \, \mu_2+\kappa_1-\kappa+1
 \end{array}\!\bigg\vert\,  q\, , \, q^{\mu_2+\kappa_2+1}\!\bigg).
\end{multline*}
The direct comparison of the latter to the former \eqref{3f2-al} leads to the very important symmetry property for the coefficients given by
\begin{equation}\label{eq_simetria}
\braket{\kappa_1\mu_1,\kappa_2\mu_2}{\kappa\mu}_q
=(-1)^{\kappa-\kappa_1-\kappa_2-1}\braket{\kappa_2\mu_2,\kappa_1\mu_1}{\kappa\mu}_{1/q}.
\end{equation}
Note that this relation could also be obtained in a similar way from the formula \eqref{3f2-smi} by applying again the transformation \eqref{142q} but with the choice
$$
n=\kappa-\kappa_1-\kappa_2-1,\ a=\kappa+\kappa_1+\kappa_2+2,\ b=\kappa_2+1-\mu_2,\ d=\kappa_1+\kappa_2+2-\mu\ \text{and}\ e=2\kappa_2+2
$$
this time. In this case the relation \eqref{3f2-smi} becomes
\begin{multline}\label{3f2-smi-sim}
\braket{\kappa_1\mu_1,\kappa_2\mu_2}{\kappa\mu}_q=
\frac{[\mu-\kappa_1-\kappa_2-2]_q!}{[2\kappa_2+1]_q!}
\sqrt{\frac{[2\kappa+1]_q[\kappa-\kappa_2+\kappa_1]_q![\kappa+\kappa_1+\kappa_2+1]_q!}{[\kappa+\kappa_2-\kappa_1]_q![\kappa-\kappa_1-\kappa_2-1]_!}} \\
\times q^{(\kappa+\kappa_1+\kappa_2+2)(\kappa-\kappa_1-\kappa_2-1)/2}\sqrt{\frac{[\mu_1+\kappa_1]_q![\mu_2+\kappa_2]_q!}{[\mu+\kappa]_q![\mu-\kappa-1]_q![\mu_1-\kappa_1-1]_q![\mu_2-\kappa_2-1]}} \\
\times q^{-\mu_2(\kappa+1)+\mu(\kappa_2+1)} 
{_3}F_2 \bigg(\!\begin{array}{c} -\kappa+\kappa_1+\kappa_2+1 \, , \, \kappa+\kappa_1+\kappa_2+2 \, ,\, \kappa_1+1-\mu_1 \\
2\kappa_1+2 \, , \, \kappa_1+\kappa_2+2-\mu
\end{array}\!\bigg\vert\,  q\, , \, q^{-\mu_2-\kappa_2-1}\!\bigg).
\end{multline}

\subsubsection{Some special values} Along the same lines as the previous passage, it is possible to determine some special values of the Clebsch-Gordan coefficients in a clever way. For instance, the one corresponding to the minimum weight vector $\ket{\kappa_2\,\kappa_2+1}$ could be obtained by the expression \eqref{3f2-smi} taking $\mu_2=\kappa_2+1$, then
\begin{multline*}
\braket{\kappa_1\mu_1,\kappa_2\,\kappa_2+1}{\kappa \mu}_q=
(-1)^{\kappa-\kappa_1-\kappa_2-1}\sqrt{\frac{[2\kappa+1]_q[\kappa-\kappa_1+\kappa_2]_q![\kappa+\kappa_1+\kappa_2+1]_q!}{[\kappa+\kappa_1-\kappa_2]_q![\kappa-\kappa_1-\kappa_2-1]_q![2\kappa_2+1]_q!}} \\
\times q^{(\kappa_1(\kappa_1+1)+\kappa_2(\kappa_2+1)-\kappa(\kappa+1))/2}
\sqrt{\frac{[\mu_1+\kappa_1]_q![\mu_1-\kappa_1-1]_q!}{[\mu+\kappa]_q![\mu-\kappa-1]_q!}}
q^{ \mu_1(\kappa-\kappa_1)}.
\end{multline*}
From here, by means of the symmetry property \eqref{eq_simetria}, the one corresponding to the minimum weight vector $\ket{\kappa_1\,\kappa_1+1}$ related to the first representation is
\begin{multline*}
\braket{\kappa_1\,\kappa_1+1,\kappa_2\mu_2}{\kappa \mu}_q=
\sqrt{\frac{[2\kappa+1]_q[\kappa+\kappa_1-\kappa_2]_q![\kappa+\kappa_1+\kappa_2+1]_q!}{[\kappa-\kappa_1+\kappa_2]_q![\kappa-\kappa_1-\kappa_2-1]_q![2\kappa_1+1]_q!}} \\
\times q^{ (\kappa(\kappa+1)-\kappa_1(\kappa_1+1)-\kappa_2(\kappa_2+1))/2}
\sqrt{\frac{[\mu_2+\kappa_2]_q![\mu_2-\kappa_2-1]_q!}{[\mu+\kappa]_q![\mu-\kappa-1]_q!}}
q^{ \mu_2(\kappa_2-\kappa)}.
\end{multline*}

On its behalf, the coefficient associated with the minimum weight vector $\ket{\kappa_1,\kappa_2;\kappa\,\kappa+1}$ could be computed in analogous way by means of \eqref{3f2-smi} again, but taking $\mu=\kappa+1$. In that case
\begin{multline*}
\braket{\kappa_1\mu_1,\kappa_2\mu_2}{\kappa\kappa+1}_q=
(-1)^{\mu_1-\kappa_1-1}\sqrt{\frac{[\kappa\!-\!\kappa_1\!+\!\kappa_2]_q![\kappa\!+\!\kappa_1\!-\!\kappa_2]_q![\kappa\!+\!\kappa_1\!+\!\kappa_2\!+\!1]_q![\kappa\!-\!\kappa_1\!-\!\kappa_2\!-\!1]_q!}
{[\mu_2+\kappa_2]_q![\mu_2-\kappa_2-1]_q![\mu_1-\kappa_1-1]_q![\mu_2-\kappa_2-1]_q![2\kappa]_q!}} \\
\times q^{(\kappa_1(\kappa_1+1)-\kappa_2(\kappa_2+1)-\kappa(\kappa+1))/2+\mu_1(\kappa-\kappa_1-\kappa_2-1)-\mu_2(\kappa_1+\kappa_2+2)}.
\end{multline*}

Finally, considering the minimum value of $\kappa=\kappa_1+\kappa_2+1$ in the formula \eqref{3f2-al}, we get the expression
\begin{multline*}
\braket{\kappa_1\mu_1,\kappa_2\mu_2}{\kappa_1+\kappa_2+1\, \mu}_q
=\sqrt{\frac{[2\kappa_1+2\kappa_2+3]_q!}{[2\kappa_1+1]_q![2\kappa_2+1]_q!}} \\
\times\sqrt{\frac{[\mu-\kappa_1-\kappa_2-2]_q![\mu_1+\kappa_1]_q![\mu_2+\kappa_2]_q!}{[\mu+\kappa_1+\kappa_2+1]_q![\mu_1-\kappa_1-1]_q![\mu_2-\kappa_2-1]_q!}}q^{\mu_1(\kappa_2+1)-\mu_2(\kappa_1+1)},
\end{multline*}
which leads to 
\begin{multline*}
\braket{\kappa_1\mu_1,\kappa_2\mu_2}{\kappa_1+\kappa_2+1\, \kappa_1+\kappa_2+2}_q
=\sqrt{\frac{[\mu_1+\kappa_1]_q![\mu_2+\kappa_2]_q!}{[2\kappa_1+1]_q![2\kappa_2+1]_q![\mu_1-\kappa_1-1]_q![\mu_2-\kappa_2-1]_q!}} \\
\times q^{\mu_1(\kappa_2+1)-\mu_2(\kappa_1+1)}.
\end{multline*}
This enable us to compute the normalization
$$
\braket{\kappa_1\, \kappa_1+1,\kappa_2\, \kappa_2+1}{\kappa_1+\kappa_2+1\, \kappa_1+\kappa_2+2}_q=1.
$$
 
\subsubsection{Connection with the $q$-Hahn polynomials.} Next, we stablish the connection of the Clebsch-Gordan coefficients with the $q$-Hahn polynomials, whose main properties
are stated in Appendix~\ref{AA}.

Comparing expression \eqref{3f2-smi-sim} with the $q$-hypergeometric representation \eqref{q-hahn1-hyp} of
the $q$-Hahn polynomials $h_n^{\alpha,\beta}$  and the one \eqref{3f2-smi} with the $q$-hypergeometric representation \eqref{q-hahn2-hyp} of the dual $q$-Hanh polynomials $W^{(c)}_{n}(s,a,b)$, we obtain the relations
\begin{gather} \label{q-hahn1}
\braket{\kappa_1\mu_1,\kappa_2\mu_2}{\kappa\mu}_q=\sqrt{\frac{\rho(s) q^{2s-1}}{d_n^2}} 
h_n^{\alpha,\beta}(s,N)_{q}, \\ \nonumber
 s=\mu_1-\kappa_1-1,\quad n=\kappa-\kappa_1-\kappa_2-1, \quad N=\mu-\kappa_1-\kappa_2-1,\quad\alpha=2\kappa_2+1,\quad \beta=2\kappa_1+1. 
\end{gather}
and 
\begin{gather}
\label{q-hahn2}
\braket{\kappa_1\mu_1,\kappa_2\mu_2}{\kappa\mu}_{q}=(-1)^{j-j_1-m_2} \sqrt{\frac{\rho(s)[2s+1]_q}{d_n^2}} W_{n}^{(c)}(s,a,b)_q,
\\ \nonumber
n=\mu_2-\kappa_2-1, \quad s=\kappa, \quad a=\kappa_1+\kappa_2+1, \quad b=\mu, \quad c=\kappa_2-\kappa_1,
\end{gather}
respectively. Above, $\rho$ and $d_n$ denote the weight function and the norm of the corresponding polynomials, respectively.

The expressions \eqref{q-hahn1} and \eqref{q-hahn2} can be used to obtain several relations for the Clebsch-Gordan coefficients by means of the ones stated for the $q$-Hahn polynomials. For example, from the orthogonality relations \eqref{ort-rel-pol} 
of the $q$-Hahn and dual $q$-Hahn polynomials the following orthogonality conditions follow
$$
\sum_{\mu_1=\kappa_1+1}^{\mu-\kappa_1-\kappa_2-1} \braket{\kappa_1\mu_1,\kappa_2\mu_2}{\kappa\mu}_{q}\braket{\kappa_1\mu_1,\kappa_2\mu_2}{\kappa'\mu}_{q}=
\delta_{\kappa,\kappa'},
$$
and
$$
\sum_{\kappa=\kappa_1+\kappa_2+1}^{\mu-1} \braket{\kappa_1\mu_1,\kappa_2\mu_2'}{\kappa\mu}_{q}\braket{\kappa_1\mu_1,\kappa_2\mu_2}{\kappa\mu}_{q}=
\delta_{\mu_2,\mu_2'},
$$
respectively. Notice that if we use the symmetry property \eqref{eq_simetria} in the first orthogonality relation it becomes
$$
\sum_{\mu_2=\kappa_2+1}^{\mu-\kappa_1-\kappa_2-1} \braket{\kappa_1\mu_1,\kappa_2\mu_2}{\kappa\mu}_{q}\braket{\kappa_1\mu_1,\kappa_2\mu_2}{\kappa'\mu}_{q}=
\delta_{\kappa,\kappa'}.
$$
Furthermore, the second order linear $q$-difference equation \eqref{dif-eq-pol} and the three-term recurrence relation \eqref{ttrr-pol}, as well as the raising and lowering relations \eqref{low-rel} and \eqref{rai-rel}, respectively, lead to several recurrence relations for the Clebsch-Gordan coefficients $\braket{\kappa_1\mu_1,\kappa_2\mu_2}{\kappa\mu}_q$. For example, the $q$-difference equation of the $q$-Hahn polynomials gives a relation involving the Clebsch-Gordan coefficients $\braket{\kappa_1\mu_1,\kappa_2\mu_2}{\kappa\mu}_q$, $\braket{\kappa_1\,\mu_1+1,\kappa_2\mu_2}{\kappa\mu}_q$, and $\braket{\kappa_1\,\mu_1-1,\kappa_2\mu_2}{\kappa\mu}_q$; whereas for the dual $q$-Hahn the involved ones are (which coincide with the ones from the three-term recurrence relation of the $q$-Hahn) $\braket{\kappa_1\mu_1,\kappa_2\mu_2}{\kappa\mu}_q$, $\braket{\kappa_1\mu_1,\kappa_2\mu_2}{\kappa\,\mu+1}_q$, and $\braket{\kappa_1\mu_1,\kappa_2\mu_2}{\kappa\,\mu-1}_q$. The three-term recurrence relation of the  $q$-Hahn polynomials relates the Clebsch-Gordan coefficients $\braket{\kappa_1\mu_1,\kappa_2\mu_2}{\kappa\mu}_q$, $\braket{\kappa_1\mu_1,\kappa_2\,\mu_2-1}{\kappa\mu}_q$, and $\braket{\kappa_1\mu_1,\kappa_2\,\mu_2+1}{\kappa\mu}_q$. Finally, the lowering and raising relations for the $q$-Hahn polynomials mix the Clebsh-Gordan coefficients $\braket{\kappa_1\mu_1,\kappa_2\mu_2}{\kappa\mu}_q$, $\braket{\kappa_1\mu_1,\kappa_2\mu_2}{\kappa\,\mu+1}_q$, and $\braket{\kappa_1\,\mu_1\mp1,\kappa_2\mu_2}{\kappa\mu}_q$, whereas the ones for the dual $q$-Hahn polynomials lead to several recurrence relations for $\braket{\kappa_1\mu_1,\kappa_2\mu_2}{\kappa\mu}_q$, $\braket{\kappa_1\mu_1,\kappa_2\,\mu_2+1}{\kappa\mu}_q$, and $\braket{\kappa_1\mu_1,\kappa_2\mu_2}{\kappa\,\mu\mp1}_q$. We leave the detailed calculations to the interested reader.

\section{Clebsch-Gordan coefficients for mixed representations of $su_q(1,1)$}
In this section we consider again the direct product of two representations $D^{\kappa+}$ and $D^{j}$ of the $su_q(1,1)$ algebra, but this time $D^{\kappa+}$ is an irreducible  and unitary (infinite dimensional) representation whereas $D^{j}$ is an non-irreducible (finite-dimensional) representation. This situation has been previously considered in \cite{ran-mt} and \cite{aiz} independently; moreover, the results obtained in the former, which are based on the theory of projection operators, were partially published in the monograph~\cite{gromov}. The key point is the fact that the generators of the $D^{\kappa+}$ representation are, as we already known, the operators $K_0$, $K_{+}$ and $K_{-}$, whereas the ones of the $D^{j}$ are the same generators of the quantum algebra $su_{q}(2)$, for which we shall use the usual notation $J_0$, $J_{+}$ and $J_{-}$, see e.g.~\cite{ran-aag}. It is worth to succinctly recalling that the action of the latter on a basis
$$
\ket{jm}, \qquad j=0,\frac{1}{2},1,\frac{3}{2},\ldots, \qquad m=-j,-j+1,\ldots,j-1,j,
$$
of the representation $D^{j}$ is
\begin{eqnarray*}
&J_{0}^{r}\ket{jm}=m^{r}\ket{jm}, \\[7pt]
&J_{+}^{r}\ket{jm}=\sqrt{\dfrac{[j-m]_q!\qn{j+m+r}!}{\qn{j+m}!\qn{j-m-r}!}}\ket{j\,m+r}, \\ [7pt]
&J_{-}^{r}\ket{jm}=\sqrt{\dfrac{\qn{j+m}!\qn{j-m+r}!}{\qn{j-m}!\qn{j+m-r}!}}\ket{j\,m-r},
\end{eqnarray*}
where $r\in\mathbb{N}$.

As it was firstly showed in \cite{ran-mt} and \cite{aiz}, the direct product of two representations of the aforementioned kinds could be expressed by means of the direct sum
\begin{equation}\label{direcsumsu11y2}
D^{\kappa+}\otimes D^{j}=\bigoplus_{\kappa'=\max\{\kappa-j,j-\kappa\}}^{\kappa+j}D^{\kappa'+},
\end{equation}
so the same reasoning carried out in Section \ref{sec:cgcReps} applies here.
However, prior to proceeding, it is necessary to point out that the theoretical minimum value for $\kappa'$ should be $\max\{0,\kappa-j\}$ as we will shown 
later on.

\subsection{Generators of the direct product representation.}
Let us consider a basis for the representation $D^{\kappa+}$, given by
$$
\ket{\kappa\mu}, \qquad \kappa=0,\frac{1}{2},1,\frac{3}{2},\ldots, \qquad \mu=\kappa+1,\kappa+2,\ldots,
$$
and the aforementioned basis $\ket{jm}$ for $D^{j}$. Then, the direct product representation has $\ket{\kappa\mu}\ket{jm}$ as a basis and it is generated by the operators
$$
K'_0(12)=K_0\otimes 1+1\otimes J_0 
\quad\text{and}\quad
K'_\pm(12)=K_\pm\otimes q^{J_0}\pm q^{-K_0}\otimes J_\pm,
$$
but, for convenience, we will use the notation
$$
K'_0(12)=K_0+J_0
\quad\text{and}\quad
K'_\pm(12)=K_\pm q^{J_0}\pm q^{-K_0}J_\pm.
$$

Few comments are in order here. First, and in the same way as in the case of the definitions of the generators for the direct product of two irreducible representations (vid.~\S \ref{subsec-CGCIR}), the factors $q^{K_0}$ and $q^{-J_0}$ in the generators $K_{\pm}(12)$ are required due to the lack of linearity of the quantum number, and they satisfy the properties
\begin{align*}
&q^{aJ_0}J_0=J_{0}q^{aJ_0}, & &q^{aJ_0}J_\pm=J_\pm q^{a(J_0\pm1)}, \\
&q^{aK_0}K_0=K_{0}q^{aK_0},& &q^{aK_0}K_\pm=K_\pm q^{a(K_0\pm1)}.
\end{align*}
On the other hand, if we define the generators
$$
\tilde{K}'_\pm(12)=K_\pm\otimes q^{-J_0}\pm q^{K_0}\otimes J_\pm.
$$
instead of the $K'_{\pm}(12)$, then $K'_{0}(12)$, $\tilde{K}'_{+}(12)$ and $\tilde{K}'_{-}(12)$ satisfy the same properties providing the minor change $q^{-1}$ in place of $q$ and, consequently, the same kind of results we shall showed are obtained.

The commutation relations for the generators of the direct product representation are
$$
\cm{K'_0(12)}{K'_\pm(12)}=\pm K'_\pm(12)
\quad\text{and}\quad
\cm{K'_{+}(12)}{K'_{-}(12)}=-\qn{2K'_0(12)},
$$
which are proved by a straightforward calculation, as well as its adjointness properties
$$
(K'_0)^\dag(12)=K'_0(12) \quad\text{and}\quad (K'_\pm)^\dag(12)=K'_\mp(12).
$$
On its behalf, the action of the generators on the basis is given by
$$
K'_0(12)\ket{\kappa\mu}\ket{jm}=(\mu+m)\ket{\kappa\mu}\ket{jm},
$$
\begin{multline*}
K'_{+}(12)\ket{\kappa\mu}\ket{jm}
=\sqrt{[\mu-\kappa]_q[\mu+\kappa+1]_q}q^{m}\ket{\kappa\,\mu+1}\ket{jm} \\
+q^{-\mu}\sqrt{[j-m]_q[j+m+1]_q}\ket{\kappa\mu}\ket{j\,m+1},
\end{multline*}
\begin{multline*}
K'_{-}(12)\ket{\kappa\mu}\ket{jm}
=\sqrt{[\mu+\kappa]_q[\mu-\kappa-1]_q}q^{m}\ket{\kappa\,\mu-1}\ket{jm} \\
-q^{-\mu}\sqrt{[j+m]_q[j-m+1]_q}\ket{\kappa\mu}\ket{j\,m-1},
\end{multline*}
or, by means of the matrix elements, in the condensed way for the last two
$$
K'_\pm(12)\ket{\kappa\mu}\ket{jm}
=\mel{\kappa\,\mu\pm1}{K_\pm(1)}{\kappa\mu}q^{m}\ket{\kappa\,\mu\pm1}\ket{jm}
\pm q^{-\mu}\mel{j\,m\pm1}{J_\pm(2)}{jm}\ket{\kappa\mu}\ket{j\,m\pm1}.
$$
Regarding the composition of the generators acting on the basis, its is clear by an induction procedure that
$$
{K'_0}^r(12)\ket{\kappa\mu}\ket{jm}=(\mu+m)^r\ket{\kappa\mu}\ket{jm},
$$
and
\begin{align}
{K'_\pm}^r(12)&=\sum_{\ell=0}^{r}\frac{(\pm1)^{r-\ell}\qn{r}!}{\qn{\ell}!\qn{r-\ell}!}K_\pm^\ell(1)J_\pm^{r-\ell}(2)q^{\ell J_0(2)-(r-\ell)K_0(1)} \label{eq11bis_BinExp}.
\end{align}

\subsection{Clebsch-Gordan coefficients.}
After the statement of the preliminary facts about the generators of the direct product representations, we are in position to compute its Clebsch-Gordan coefficients $\braket{\kappa\mu,jm}{\kappa'\mu'}_q$ in the expansion~\eqref{jm-jimi}.  

Since its expression given by \eqref{eq_GenCG} is totally independent of the choice of $\bar{\mu}_1$ and $\bar{\mu}_2$, which is in fact the key point, we are able to take $\bar{\mu}_1=\kappa+1$ and $\bar{\mu}_2=\kappa'-\kappa$, which implies $m'=m'_1+m'_2=j+1$. The reasons for setting $m'_1$ the minimum value of $\mu$ and $m'_2$ such that $m'=j+1$ will be clear soon. Therefore
\begin{equation}\label{eq_CG}
\braket{\kappa\mu,jm}{\kappa'\mu'}_q=\frac{\bra{\kappa\mu}\bra{jm}P^{\kappa'+}_{\mu',\kappa'+1}\ket{\kappa\,\kappa+1}\ket{j\,\kappa'-\kappa}}{\|P^{\kappa+}_{\mu',\kappa'+1}\ket{\kappa\,\kappa+1}\ket{j\,j-\kappa}\|},
\end{equation}
where  $\kappa'\geq\max\{0,\kappa-j\}$, which is imposed by our choice since $-j\leq \bar{\mu}_{2}=\kappa'-\kappa\leq j$.
 
The main tool to do the calculation of the numerator in \eqref{eq_CG} is the generalized projection operator  showed in \eqref{su11_proj-gen}.  Here, the advantage of our selection in the parameters such that $\bar{\mu}=\kappa'+1$ appears clearly, because its expression is the quite simply
 \begin{align*}
P^{\kappa'+}_{\mu',\kappa'+1}
&=\sqrt{\frac{\qn{2\kappa'+1}!}{\qn{\mu'-\kappa'-1}!\qn{\mu'+\kappa'}!}}{K'}_{+}^{\mu'-\kappa'-1}(12)P^{\kappa'+}_{\kappa'+1\,\kappa'+1} \\
&=\sqrt{\frac{\qn{2\kappa'+1}!}{\qn{\mu'-\kappa'-1}!\qn{\mu'+\kappa'}!}}\sum_{r=0}^{\infty}\frac{(-1)^{r}\qn{2\kappa'-r}!}{\qn{r}!\qn{2\kappa'}!}
{K'}_{+}^{\mu'-\kappa'-1+r}(12){K'}_{-}^{r}(12).
\end{align*}
Then, the numerator is given by
\begin{multline}\label{eq11b_NumCGHExp}
\bra{\kappa\mu}\bra{jm} \sqrt{\frac{\qn{2\kappa'+1}!}{\qn{\mu'-\kappa'-1}!\qn{\mu'+\kappa'}!}} \\
\times \sum_{r=0}^{\kappa'}\frac{(-1)^{r}\qn{2\kappa'-r}!}{\qn{r}!\qn{2\kappa'}}{K'_+}^{\mu'-\kappa'-1+r}(12){K'_-}^{r}(12)
\ket{\kappa\,\kappa+1}\ket{j\,\kappa'-\kappa}.
\end{multline}
On the one hand, by means of the binomial expansion \eqref{eq11bis_BinExp}, we get
\begin{multline*}
{K'_-}^{r}(12)\ket{\kappa\,\kappa+1}\ket{j\,\kappa'-\kappa}
=
\sum_{\ell=0}^{r}\frac{(-1)^{r-\ell}\qn{r}!}{\qn{\ell}!\qn{r-\ell}!}\\
\times 
\sqrt{\frac{\qn{2\kappa+1}!\qn{j-\kappa+\kappa'}!\qn{j+\kappa-\kappa'+r-\ell}!}{\qn{2\kappa+1-\ell}!\qn{-\ell}!\qn{j+\kappa-\kappa'}!\qn{j-\kappa+\kappa'-r+\ell}!}} \\
\times q^{\ell(\kappa'-\kappa)-(r-\ell)(\kappa+1)}\ket{\kappa\,\kappa+1-\ell}\ket{j\,\kappa'-\kappa-r+\ell}
\end{multline*}
but, thanks to our previous selection for $\bar{\mu}_1$ as the minimum value $\kappa+1$, the only possible addend in the sum is the one corresponding with $\ell=0$, which implies
$$
{K'_-}^{r}(12)\ket{\kappa\,\kappa+1}\ket{j\,\kappa'-\kappa}
=(-1)^{r}\sqrt{\frac{\qn{j-\kappa+\kappa'}!\qn{j+\kappa-\kappa'+r}!}{\qn{j+\kappa-\kappa'}!\qn{j-\kappa+\kappa'-r}!}}q^{-(\kappa+1)r}\ket{\kappa\,\kappa+1}\ket{j\,\kappa'-\kappa-r}.
$$
Therefore, the numerator \eqref{eq11b_NumCGHExp} becomes
\begin{multline}\label{eq11b_BinExp2}
\bra{\kappa\mu}\bra{jm}\sqrt{\frac{\qn{2\kappa'+1}!\qn{j-\kappa+\kappa'}!}{\qn{\mu'+\kappa'}!\qn{\mu'-\kappa'-1}!\qn{j+\kappa-\kappa}!}}
\sum_{r=0}^{j-\kappa+\kappa'}\frac{\qn{2\kappa'-r}!}{\qn{r}!\qn{2\kappa'}!}\sqrt{\frac{\qn{j+\kappa-\kappa'+r}!}{\qn{j-\kappa+\kappa'-r}!}} \\[5pt]
\times  q^{-(\kappa+1)r}{K'_+}^{\mu'-\kappa'-1+r}(12)\ket{\kappa\,\kappa+1}\ket{j\,\kappa'-\kappa-r}.
\end{multline}
On the other hand, using again the binomial expansion \eqref{eq11bis_BinExp}, we get
\begin{multline*}
{K'_+}^{\mu'-\kappa'-1+r}(12)\ket{\kappa\,\kappa+1}\ket{j\,\kappa'-\kappa-r}
=q^{-(\mu'-\kappa'-1+r)(\kappa+1)}\qn{\mu'-\kappa'-1+r}! \\[2pt]
\times\sum_{\ell=\max\{0,\mu'-\kappa-j-1\}}^{\mu'-\kappa'-1+r}\frac{q^{(\kappa'+1-r)\ell}}{\qn{\mu'-\kappa'-1+r-\ell}!}
\sqrt{\frac{\qn{2\kappa+1+\ell}!\qn{j+\kappa-\kappa'+r}!\qn{j-\kappa+\mu'-1-\ell}}{\qn{\ell}!\qn{2\kappa+1}!\qn{j-\kappa+\kappa'-r}!\qn{j+\kappa-\mu'+1+\ell}!}} \\
\times\ket{\kappa\,\kappa+1+\ell}\ket{j\, \mu'-\kappa-1-\ell}.
\end{multline*}
Therefore, the expression \eqref{eq11b_BinExp2} turns into
\begin{multline*}
\sqrt{\frac{\qn{2\kappa'+1}\qn{j-\kappa+\kappa'}!}{\qn{2\kappa'}!\qn{2\kappa+1}!\qn{j+\kappa-\kappa'}!\qn{\mu'+\kappa'}!\qn{\mu'-\kappa'-1}!}}
q^{-(\mu-\kappa'-1)(\kappa+1)} \\
\times \sum_{r=0}^{j-\kappa+\kappa'}\frac{\qn{2\kappa'-r}!\qn{j+\kappa-\kappa'+r}!\qn{\mu'-\kappa'-1+r}!}{\qn{r}!\qn{j-\kappa+\kappa'-r}!}{q^{-2(\kappa+1)r}} \\
\times\sum_{\ell=\max\{0,\mu'-\kappa-j-1\}}^{\mu'-\kappa'-1+r}\frac{q^{(\kappa'+1-r)\ell}}{\qn{\mu'-\kappa'-1+r-\ell}!}\sqrt{\frac{\qn{2\kappa+1+\ell}!\qn{j-\kappa+\mu'-1-\ell}!}{\qn{\ell}!\qn{j+\kappa-\mu'+1+\ell}!}} \\
\times\braket{\kappa\mu}{\kappa\,\kappa+1+\ell}\braket{jm}{j\,\mu'-\kappa-1-\ell}.
\end{multline*}
However, by means of the orthogonality of the basis, there is again only one addend in the inner sum which corresponds with $\ell=\mu-\kappa-1$. Therefore, after a rearrangement, the numerator of the Clebsch-Gordan coefficient is
\begin{multline}\label{eq11b_NumCGFinal}
\bra{\kappa\mu}\bra{jm}P^{\kappa'+}_{\mu',\kappa'+1}\ket{\kappa\,\kappa+1}\ket{j\,\kappa'-\kappa}= \\
\times \sqrt{\frac{\qn{2\kappa'+1}\qn{j-\kappa+\kappa'}!\qn{\mu+\kappa}!\qn{j+m}!}{\qn{2\kappa'}!\qn{2\kappa+1}!\qn{\kappa+j-\kappa'}!\qn{\mu'+\kappa'}!\qn{\mu'-\kappa'-1}!\qn{\mu-\kappa-1}!\qn{j-m}!}} q^{\mu(\kappa'+1)-\mu'(\kappa+1)} \\
\times \sum_{ {r=\max\{0,\kappa'-\kappa-m\}}}^{j-\kappa+\kappa'}\frac{\qn{2\kappa'-r}!\qn{\kappa+j-\kappa'+r}!\qn{\mu'-\kappa'-1+r}!}{\qn{r}!\qn{j-\kappa+\kappa'-r}!\qn{m+\kappa-\kappa'+r}!}q^{-(\mu+\kappa+1)r},
\end{multline}
where the $\max\{0,\kappa'-\kappa-m\}$ follows from the condition  $\ell=\mu-\kappa-1\leq \mu'-\kappa'-1+r$. 
It is worth noting that the upper limit in the sum is always greater than or equal to zero, regardless of whether $j>\kappa$ or $\kappa\geq j$, and it is always greater than or equal to the lower limit of the summation. Notice also that, if $\kappa'-\kappa-m>0$, and taking into account that 
$$
\frac{1}{\qn{m+\kappa-\kappa'+r}!}=(-1)^{\kappa'-\kappa-m}\frac{(-r|q)_{\kappa'-\kappa-m}}{\qn{r}!},
$$ 
the lower index of the sum in the last expression can be taken equal zero because all the extra terms will vanish. In fact, the sum can be taken over the non-negative integers because, regardless of which upper limit is larger, all the additional terms vanish as well.

The denominator of the Clebsch–Gordan coefficient in expression \eqref{eq_CG}, in turn, can be evaluated using the identity
$$
\|P^{\kappa'+}_{\mu',\kappa'+1}\ket{\kappa\,\kappa+1}\ket{j\,\kappa'-\kappa}\|^2
=\bra{\kappa\,\kappa+1}\bra{j\,\kappa'-\kappa}P^{\kappa'+}_{\mu',\kappa'+1}\ket{\kappa\,\kappa+1}\ket{j\,\kappa'-\kappa}
$$
and the value of the numerator \eqref{eq11b_NumCGFinal} with the choice of parameters $\mu=\kappa+1$ and $m=\kappa'-\kappa$. Then
\begin{multline*}
\|P^{\kappa'+}_{\mu',\kappa'+1}\ket{\kappa\,\kappa+1}\ket{j\,\kappa'-\kappa}\|^2
=\frac{\qn{j-\kappa+\kappa'}!}{\qn{2\kappa'}!\qn{\kappa+j-\kappa'}!} 
\sum_{r=0}^{\infty}\frac{\qn{2\kappa'-r}!\qn{\kappa+j-\kappa'+r}!}{\qn{r}!\qn{j-\kappa+\kappa'-r}!}q^{-2(\kappa+1)r},
\end{multline*}
but the series above is actually finite and its sum is given by
\begin{multline*}
\sum_{r=0}^{\infty}\frac{\qn{2\kappa'-r}!\qn{\kappa+j-\kappa'+r}!}{\qn{r}!\qn{j-\kappa+\kappa'-r}!}q^{-2(\kappa+1)r}
=\frac{\qn{\kappa+j+\kappa'+1}!\qn{\kappa+j-\kappa'}!\qn{\kappa-j+\kappa'}!}{\qn{2\kappa+1}!\qn{j-\kappa+\kappa'}!} \\
\times q^{-(\kappa+j-\kappa'+1)(j_2-\kappa+\kappa')},
\end{multline*}
where we have used formula \eqref{suma3equiv}. So the expression for the denominator is
\begin{equation*}
\|P^{\kappa'+}_{\mu,\kappa'+1}\ket{\kappa\,\kappa+1}\ket{j\,\kappa'-\kappa}\|
=\sqrt{\frac{\qn{\kappa+j+\kappa'+1}!\qn{\kappa-j+\kappa'}!}{\qn{2\kappa'}!\qn{2\kappa+1}!}}q^{-(\kappa+j-\kappa'+1)(j-\kappa+\kappa')/2}.
\end{equation*}

Therefore, the Clebsch-Gordan coefficients for mixed representations of the $su_q(1,1)$ algebra are given by
\begin{multline}\label{eq_cgMixed-a}
\braket{\kappa\mu,jm}{\kappa'\mu'}_q=
\sqrt{\frac{\qn{2\kappa'+1}\qn{j-\kappa+\kappa'}!}{\qn{\kappa+j+\kappa'+1}!\qn{\kappa-j+\kappa'}!\qn{\kappa+j-\kappa'}!}}q^{(\kappa+j-\kappa'+1)(j-\kappa+\kappa')/2} \\
\times \sqrt{\frac{\qn{\mu+\kappa}!\qn{j+m}!}{\qn{\mu'+\kappa'}!\qn{\mu'-\kappa'-1}!\qn{\mu-\kappa-1}!\qn{j-m}!}}
q^{\mu(\kappa'+1)-\mu'(\kappa+1)} \\
\times \sum_{r=0}^{\infty}\frac{\qn{2\kappa'-r}!\qn{\kappa+j-\kappa'+r}!\qn{\mu'-\kappa'-1+r}!}{\qn{r}!\qn{j-\kappa+\kappa'-r}!\qn{m+\kappa-\kappa'+r}!}q^{-(\mu+\kappa+1)r}.
\end{multline}
By performing the change of variables  $r=j-\kappa+\kappa'-s$ 
in the above expression, we obtain the following equivalent form
\begin{multline}\label{eq_cgMixed-b}
\braket{\kappa\mu,jm}{\kappa'\mu'}_q=
\sqrt{\frac{\qn{2\kappa'+1}\qn{j-\kappa+\kappa'}!}{\qn{\kappa+j+\kappa'+1}!\qn{\kappa-j+\kappa'}!\qn{\kappa+j-\kappa'}!}}q^{-(\kappa-j+\kappa'+1)(j-\kappa+\kappa')/2} \\
\times \sqrt{\frac{\qn{\mu+\kappa}!\qn{j+m}!}{\qn{\mu'+\kappa'}!\qn{\mu'-\kappa'-1}!\qn{\mu-\kappa-1}!\qn{j-m}!}}
q^{\mu(\kappa-j+1)-\mu'(\kappa+1)} \\
\times \sum_{r=0}^{\infty}\frac{\qn{2j-r}!\qn{\kappa-j+\kappa'+r}!\qn{\mu'-\kappa+j-1-r}!}{\qn{r}!\qn{j-\kappa+\kappa'-r}!\qn{j+m-r}!}q^{(\mu+\kappa+1)r},
\end{multline}
where we have recovered the original variable~$r$. Moreover, since all of the argument of the involved $q$-factorials in \eqref{eq_cgMixed-a} and \eqref{eq_cgMixed-b} should be positive, we have $\max\{j-\kappa,\kappa-j\}\leq \kappa' \leq \kappa+j$. This justify the lower and upper indices given in the aforementioned direct sum~\eqref{direcsumsu11y2}, that is in agreement with the results in~\cite{aiz,ran-mt}.
 
\subsubsection{Representations as a terminating $q$-hypergeometric series}
The expression \eqref{eq_cgMixed-a} for the Clebsch–Gordan coefficients can be reformulated in terms of the symmetric terminating $q$-hypergeometric function
${_3}F_{2}$ by applying the identities in \eqref{q-fac-poc}, yielding
\begin{multline}\label{eq_cgMixed3f2-a}
\braket{\kappa\mu,jm}{\kappa'\mu'}_q=
\sqrt{\frac{\qn{2\kappa'}!\qn{2\kappa'+1}!\qn{\kappa+j-\kappa'}!}{\qn{\kappa+j+\kappa'+1}!\qn{\kappa-j+\kappa'}!\qn{j-\kappa+\kappa'}!}}
q^{(\kappa+j-\kappa'+1)(j-\kappa+\kappa')/2} \\
\times \frac{1}{\qn{m+\kappa-\kappa'}!} 
\sqrt{\frac{\qn{\mu'-\kappa'-1}!\qn{\mu+\kappa}!\qn{j+m}!}{\qn{\mu'+\kappa'}!\qn{\mu-\kappa-1}!\qn{j-m}!}}q^{\mu(\kappa'+1)-\mu'(\kappa+1)} \\
\times  {_3}F_2 \bigg(\!\begin{array}{c}  \kappa-j-\kappa' \, , \, \mu'-\kappa' \, ,\, \kappa+j-\kappa'+1 \\
-2\kappa' \, , \, m+\kappa-\kappa'+1
 \end{array}\!\bigg\vert\,  q\, , \, q^{-(\mu+\kappa+1)}\!\bigg).
\end{multline}
A similar reasoning can be applied to expression \eqref{eq_cgMixed-b}, leading to the formula
\begin{multline}\label{eq_cgMixed3f2-b}
\braket{\kappa\mu,jm}{\kappa'\mu'}_q=
\qn{2j}!\sqrt{\frac{\qn{2\kappa'+1}\qn{\kappa-j+\kappa'}!}{\qn{\kappa+j+\kappa'+1}!\qn{j-\kappa+\kappa'}!\qn{\kappa+j-\kappa'}!}}
q^{-(\kappa-j+\kappa'+1)(j-\kappa+\kappa')/2} \\
\times
\qn{\mu'-\kappa+j-1}!\sqrt{\frac{\qn{\mu+\kappa}!}{\qn{\mu'+\kappa'}!\qn{\mu'-\kappa'-1}!\qn{\mu-\kappa-1}!\qn{j+m}!\qn{j-m}!}}q^{\mu j-m(\kappa+1)} \\
\times
 {_3}F_2 \bigg(\!\begin{array}{c}  \kappa-j-\kappa' \, , \, -j-m \, ,\, \kappa-j+\kappa'+1 \\
-2j \, , \, \kappa-j-\mu'+1
 \end{array}\!\bigg\vert\,  q\, , \, q^{\mu+\kappa+1}\!\bigg).
\end{multline}

Furthermore, alternative formulae can be derived through suitable transformations of the symmetric $q$-hypergeometric functions. In particular, setting $n=j-\kappa+\kappa'$, $a=\mu'-\kappa'$, $b=\kappa+j-\kappa'+1$, $c=-2\kappa'$ and $e=m+\kappa-\kappa'+1$ in formula \eqref{142q}, expression \eqref{eq_cgMixed3f2-a} becomes
\begin{multline}\label{eq_cgMixed3f2-c}
\braket{\kappa\mu,jm}{\kappa'\mu'}_q
=(-1)^{j-\kappa+\kappa'}\sqrt{\frac{\qn{2\kappa'}!\qn{2\kappa'+1}!\qn{\kappa+j-\kappa'}!}{\qn{\kappa+j+\kappa'+1}!\qn{\kappa-j+\kappa'}!\qn{j-\kappa+\kappa'}!}}
 \\
\times\frac{1}{\qn{\mu-j-\kappa'-1}!}\sqrt{\frac{\qn{\mu'-\kappa'-1}!\qn{\mu+\kappa}!\qn{\mu-\kappa-1}!}{\qn{\mu'+\kappa'}!\qn{j+m}!\qn{j-m}!}}q^{-\mu j-m(j+\kappa'+1)} \\
\times q^{(\kappa+j+\kappa'+1)(j-\kappa+\kappa')/2}\,\, {_3}F_2 \bigg(\!\begin{array}{c}  \kappa-j-\kappa' \, , \, \mu'-\kappa' \, ,\, -\kappa-j-\kappa'-1 \\
-2\kappa' \, , \, \mu-j-\kappa'
 \end{array}\!\bigg\vert\,  q\, , \, q^{m-j}\!\bigg).
\end{multline}

\subsubsection{Some special values} The representations of the Clebsch-Gordan coefficients as a terminating $q$-hypergeometric series allow to compute easily some special values by means of its summation formulae.

Let us set $m=j$, its maximum value, in equation \eqref{eq_cgMixed3f2-a}. Then, by \eqref{suma3}, we obtain
\begin{multline*}
\braket{\kappa\mu,jj}{\kappa'\mu'}_q=\sqrt{\frac{\qn{2\kappa'+1}!\qn{\kappa-j+\kappa'}!\qn{2j}!\qn{\mu'-\kappa'-1}!\qn{\mu'+\kappa'}!}{\qn{\kappa+j+\kappa'+1}!\qn{\kappa+j-\kappa'}!\qn{j-\kappa+\kappa'}!\qn{\mu+\kappa}!\qn{\mu-\kappa-1}!}} \\
\times q^{\mu(\kappa'+1)-\mu'(j+\kappa'+1)+(\kappa+j+\kappa'+1)(j-\kappa+\kappa')/2}.
\end{multline*}
When $m=-j$, its minimum value, we have to use \eqref{eq_cgMixed3f2-c} to get
\begin{multline*}
\braket{\kappa\mu,j\,-j}{\kappa'\mu'}_q
=\sqrt{\frac{\qn{2\kappa'+1}\qn{2j}!\qn{\kappa-j+\kappa'}!\qn{\mu'+\kappa}!\qn{\mu-\kappa-1}!}{\qn{\kappa+j+\kappa'+1}!\qn{\kappa+j-\kappa'}!\qn{j-\kappa+\kappa'}!\qn{\mu'+\kappa'}!\qn{\mu'-\kappa'-1}!}} \\
\times q^{j(\mu'+\kappa'+1)-(\kappa+j+\kappa'+1)(j-\kappa+\kappa')/2},
\end{multline*}
where we have use \eqref{suma3} again. In the case of the minimum value of $\mu$, we take $\mu=\kappa+1$ in \eqref{eq_cgMixed3f2-a}, so
\begin{multline*}
\braket{\kappa\,\kappa+1,jm}{\kappa'\mu'}_q
=\sqrt{\frac{\qn{2\kappa'+1}!\qn{\kappa+j+\kappa'+1}!\qn{\kappa+j-\kappa'}!\qn{\kappa-j+\kappa'}!\qn{j+m}!}{\qn{2\kappa+1}!\qn{j-\kappa+\kappa'}!\qn{\mu'-\kappa'-1}!\qn{\mu'+\kappa'}!\qn{j-m}!}} \\
\times q^{-(\kappa+1)(\mu'-\kappa'-1)}.
\end{multline*}
Finally, fixing the minimum value $\mu'=\kappa'+1$ in the representation \eqref{eq_cgMixed3f2-b}, we obtain, as usual, by means of \eqref{suma3},
\begin{multline*}
\braket{\kappa\mu,jm}{\kappa'\,\kappa'+1}_q
=\sqrt{\frac{\qn{\kappa+j+\kappa'+1}!\qn{\kappa-j+\kappa'}!\qn{j-\kappa+\kappa'}!\qn{j-m}!}{\qn{\kappa+j-\kappa'}!\qn{\mu+\kappa}!\qn{\mu-\kappa-1}!\qn{j+m}!}} \\
\times
q^{\mu\kappa-m(\kappa+1)-(\kappa-j+\kappa'+1)(j-\kappa+\kappa')/2}.
\end{multline*}

The next special value among the ones which have been considered is obtained by setting the maximum value $\kappa'=\kappa+j$ in the representation \eqref{eq_cgMixed3f2-b},
$$
\braket{\kappa\mu,jm}{\kappa+j\,\mu'}_q
=\sqrt{\frac{\qn{2\kappa}!\qn{2j}!\qn{\mu'+\kappa+j}!\qn{\mu-\kappa-1}!}{\qn{\mu'-\kappa-j-1}!\qn{\mu+\kappa}!\qn{j+m}!\qn{j-m}!}}q^{-\mu j+m\kappa}.
$$
We obtain now the minimum value when the parameter $\kappa'$ is $\kappa-j>0$. Setting $\kappa'=\kappa-j$ in \eqref{eq_cgMixed3f2-a} or \eqref{eq_cgMixed3f2-b} we attain
$$
\braket{\kappa\mu,jm}{\kappa-j\,\mu'}_q=
\sqrt{\frac{\qn{2\kappa-2j+1}!\qn{2j}!\qn{\mu'-\kappa+j+1}!\qn{\mu+\kappa}!}{\qn{2\kappa+1}!\qn{\mu'+\kappa-j}!\qn{\mu-\kappa-1}!\qn{j+m}!\qn{j-m}!}}
q^{-\mu j-m(\kappa+1)},
$$
where we have taken into account that the hypergeometric function involved in the equation is, with choice, equal one. Here we have chosen the representation of the Clebsch-Gordan coefficient given by \eqref{eq_cgMixed3f2-a}, but we could lead to the same result by means of the \eqref{eq_cgMixed3f2-c} one. Note that this value is precisely the normalization when $\mu$, $m$ and $\mu'$ take the minimum possible value, respectively, that is
$$
\braket{\kappa\,\kappa+1,j\,-j}{\kappa-j\,\kappa-j+1}=1,\qquad \kappa\geq j.
$$
The case when  $\kappa'=j-\kappa>0$, which is also possible is rather more complicated. To obtain the value one can use our first formula \eqref{eq_cgMixed-a}, so
\begin{multline*} 
\braket{\kappa\mu,jm}{j-\kappa'\,\mu'}_q=
\sqrt{\frac{\qn{2j-2\kappa+1}!\qn{\mu+\kappa}!\qn{j+m}!}
{\qn{2j+1}!\qn{2\kappa}!\qn{\mu'+j-\kappa}!\qn{\mu'-j+\kappa-1}!\qn{\mu-\kappa-1}!\qn{j-m}! }} \\
\times q^{(2\kappa+1)(j-\kappa) +\mu(j-\kappa+1)-\mu'(\kappa+1)}\!\!\!  
\sum_{r=\max\{0,j-2\kappa-m\}}^{2j-2\kappa}\frac{ \qn{2\kappa+r}!\qn{\mu'-j+\kappa-1+r}!}{\qn{r}!\qn{m-j+2\kappa+r}!}
q^{-(\mu+\kappa+1)r}.
\end{multline*}
If $j-2\kappa-m>0$ it is convenient to make the change $r=j-2\kappa-m+s$, so the sum becomes 
$$
q^{-(\mu+\kappa+1)(j-2\kappa-m)}
\sum_{r= 0}^{j+m}\frac{ \qn{j-m+r}!\qn{\mu-\kappa-1+r}!}{\qn{r}!\qn{j-2\kappa-m+r}!}
q^{-(\mu+\kappa+1)r}.
$$
In both cases we were unable to get a closed form for the sum.

\subsubsection{Connection with the $q$-Hahn polynomials} 
Comparing expression \eqref{eq_cgMixed3f2-b} with the formula \eqref{q-hahn1-hyp} for the $q$-Hahn polynomials $h_n^{\alpha,\beta}$ we find 
\begin{gather}\label{eq_cgMixed-h}
\braket{\kappa\mu,jm}{\kappa'\mu'}_q=\Gamma^{\kappa,j,\kappa'}_{\mu,m,\mu'}
\,\, h_n^{\alpha,\beta}(s,N)_{1/q},
\end{gather}
where $n=\kappa'-\kappa+j$, $s=j+m$, $\alpha=\kappa-j-\kappa'$, $\beta=\kappa-j+\kappa'$, $N=2j+1$,
and 
\begin{multline*}
\Gamma^{\kappa,j,\kappa'}_{\mu,m,\mu'}=
\sqrt{\frac{\qn{2\kappa'+1}\qn{\kappa-j+\kappa'}!\qn{\mu+\kappa}!\qn{\mu'-\kappa'-1}!}{\qn{\kappa+j+\kappa'+1}!\qn{j-\kappa+\kappa'}!\qn{\kappa+j-\kappa'}! \qn{\mu'+\kappa'}!\qn{\mu-\kappa-1}!\qn{j+m}!\qn{j-m}!} } \\
\times q^{\mu j-m(\kappa+1)+(\kappa-j+1)(\kappa'-\kappa+1)} \qn{2j}!.
\end{multline*}
Notice that the polynomials on the right hand side are defined in terms of $1/q$. Moreover, in this case $\beta\leq-1$, and therefore the $q$-Hahn polynomials on the right hand side of \eqref{eq_cgMixed-h} are not orthogonal, so neither the Clebsch-Gordan coefficients, which is not a surprise since the representation $D^j$ is not an irreducible representation of the quantum algebra $su_q(1,1)$.

Comparing now the expression \eqref{eq_cgMixed3f2-b} with the formula \eqref{q-hahn2-hyp} of the dual $q$-Hanh polynomials $W^{(c)}_{n}(s,a,b)$ we obtain
\begin{gather}
\label{eq_cgMixed-dual}
\braket{\kappa\mu,jm}{\kappa'\mu'}_q= \widetilde{\Gamma}^{\kappa,j,\kappa'}_{\mu,m,\mu'} W_{n}^{(c)}(s,a,b)_q,
\end{gather}
where now $n=j+m$, $s=\kappa'$, $a=k-j$, $b=k+j+1$, $c=-\mu'$, and  
\begin{multline*}
\widetilde{\Gamma}^{\kappa,j,\kappa'}_{\mu,m,\mu'}
=\sqrt{\frac{\qn{2\kappa'+1}\qn{\kappa-j+\kappa'}!\qn{\mu+\kappa}!\qn{\mu-\kappa-1}!\qn{j+m}!\qn{j-m}!}{\qn{\kappa+j+\kappa'+1}!\qn{j-\kappa+\kappa'}!\qn{\kappa+j-\kappa'}! \qn{\mu'+\kappa'}!\qn{\mu'-\kappa'-1}!}} \\
\times q^{m\mu +2j\mu'+jm+2j^2-j+\frac{3m(m-1)}{2}-\frac{\kappa'(\kappa'+1)}{2}+\frac{\kappa(\kappa+1)}{2}}.
\end{multline*}
Moreover, since $|c| = \mu'$, the condition $|c| < a + 1$, which is required for orthogonality, is not fulfilled. Consequently the dual $q$-Hahn polynomials appearing on the right-hand side of Eq. \eqref{eq_cgMixed-dual} are also not orthogonal.

Although the involved polynomials are not orthogonal, they still satisfy the difference equation~\eqref{dif-eq-pol}. This property allow us to derive linear recurrence relations for the Clebsch–Gordan coefficients corresponding to the mixed representations. In fact, by substituting expressions \eqref{eq_cgMixed-h} and \eqref{eq_cgMixed-dual} into the difference equation \eqref{dif-eq-pol} we obtain the relations involving $\braket{\kappa\mu,j\,m\pm1}{\kappa'\,\mu'\pm1}_q$, $\braket{\kappa\mu,jm}{\kappa'\mu'}_q$ and $\braket{\kappa\mu,jm}{\kappa'\pm1\,\mu'}_q$, $\braket{\kappa\mu,jm}{\kappa'\mu'}_q$, respectively. Moreover, considering that equations \eqref{low-rel} and \eqref{rai-rel} hold for the polynomial solutions of the $q$-Nikiforov-Uvarov equation, regardless of whether they are orthogonal, we can also employ them to derive linear relations involving $\braket{\kappa\mu,jm}{\kappa'\pm1\,\mu'}_q$, $\braket{\kappa\mu,j\,m+1}{\kappa'\,\mu'+1}_q$, and $\braket{\kappa\mu,jm}{\kappa'\mu'}_q$. For clarity, some of these 
recurrences are presented in the Appendix~\ref{BB}.

\vspace*{1cm}

\paragraph{\textbf{Concluding remarks:}}
The theory of $q$-hypergeometric functions is also highly useful for studying the Clebsch-Gordan coefficients of the quantum algebra $su_q(1,1)$. As has been shown, several properties that are quite difficult to derive using standard tools from representation theory can be obtained more easily by expressing the Clebsch-Gordan coefficients in terms of the symmetric $q$-hypergeometric function ${}_3F_2$. In particular, classical results for the algebra (or group) $su(1,1)$ are recovered in the limit as $q \to 1$.
\bigskip

\paragraph{\textbf{Funding:}}
Renato \'Alvarez-Nodarse was partially supported by PID2021-124332NB-C21 (FEDER (EU) / Mi\-nis\-te\-rio de Cien\-cia e In\-no\-va\-ci\'on - A\-gen\-cia Es\-ta\-tal de In\-ves\-ti\-ga\-ci\-\'on) and FQM-415 (Jun\-ta deAn\-da\-lu\-c\'i\-a). Alberto Arenas-G\'omez was partially supported by PID2021-124332NB-C22 (FEDER (EU) / Mi\-nis\-te\-rio de Cien\-cia e In\-no\-va\-ci\'on - A\-gen\-cia Es\-ta\-tal de In\-ves\-ti\-ga\-ci\-\'on).\bigskip

\paragraph{\textbf{Conflicts of Interest:}}
The authors of this work declare that they have no conflicts of interest.

\bigskip 
\begin{appendices}
\noindent\appendix

\section{About the unitary irreducible representations of $su_q(1,1)$}\label{ap-al}
In the present appendix we exhibit a construction of the unitary irreducible representations of the quantum algebra $su_q(1,1)$ in the spirit of the one sketched out in~\cite[p.~25]{dong} for both algebras $so(3)$ and $so(2,1)$, which are isomorphic to $su(2)$ and $su(1,1)$, respectively.

The infinitesimal generators of the quantum algebra $su_q(1,1)$ are $K_{+}$, $K_{-}$, and $K_{0}$, which satisfy the following commutation relations:
$$
[K_{0},K_{+}]=K_{+},\ [K_{0},K_{-}]=K_{-}, \text{ and } [K_{+},K_{-}]=-[2K_{0}]_{q},
$$
just as the adjointness properties $(K_{+})^\dag=K_{-}$, $(K_{-})^\dag=K_{+}$, and $(K_{0})^\dag=K_0$. Here $[x]_{q}$ states for the symmetric quantun number defined for $x\in\mathbb{R}$ by
$$
[x]_{q}=\frac{q^{x}-q^{-x}}{q-q^{-1}}=\frac{\sinh\gamma x}{\sinh\gamma},
\qquad q=e^{\gamma}\in(0,1).
$$
On its behalf, the Casimir operator $C$ is given by
$$
C=-K_{+}K_{-}+[K_{0}]_{q}[K_{0}-1]_q=-K_{-}K_{+}+[K_{0}]_{q}[K_{0}+1]_{q}.
$$
It possesses the important property of self-adjointness, $C^{\dag}= C$, and it commutes with the infinitesimal generators; 
that is, $[C, K_{0}] = [C, K_{+}] = [C, K_{-}] = 0$. These properties allow us to choose a basis $\ket{\kappa\mu}$ in which both operators $K_0$ and $C$ are diagonalized, that is,
$$
K_{0}\ket{\kappa\mu}=\mu\ket{\kappa\mu} \quad\text{and}\quad C\ket{\kappa\mu}=[\kappa]_q[\kappa+1]_q\ket{\kappa\mu}.
$$
The eigenvalues, which are real due to the hermiticity of the operator, have been chosen in this form for simplicity. It is worth noting that the substitution $\kappa \mapsto -(\kappa + 1)$ leaves these eigenvalues invariant.

The action of the generators $K_{+}$ and $K_{-}$ is, by means of the commutation relations, $K_{0}K_{+}\ket{\kappa\mu}=(\mu+1)K_{+}\ket{\kappa\mu}$ and $K_{0}K_{-}\ket{\kappa\mu}=(\mu-1)K_{-}\ket{\kappa\mu}$. 
This implies that both operators are, in fact, ladder operators; that is, they raise or lower the eigenvalue of $K_0$ by one unit
$$
K_{+}\ket{\kappa\mu}=a_{\kappa\mu}\ket{\kappa\,\mu+1} \quad\text{and}\quad K_{-}\ket{\kappa\mu}=b_{\kappa\mu}\ket{\kappa\,\mu-1},
$$
where $a_{\kappa\mu}$ and $b_{\kappa\mu}$ are (unknown at the moment) numbers. Due to the ladder character of the operators, the spectrum of $K_0$ is of the form $\{\mu:=\mu_0+n:n\in\mathbb{Z}\}$ for a special ``first'' element $\mu_0$, so there are four different situations: i.~the spectrum is not upper bounded nor lower bounded; ii.~the spectrum is upper bounded; iii.~the spectrum is lower bounded; and, iv.~the spectrum is bounded. In the present work, we are only interested in the cases ii.~and iii., specially in the latter, also known as positive series, since the treatment of the former, called in contrast negative series, is completely analogous. Then, from now on, we assume that there exits an element $\mu_0$ such that $K_{-}\ket{\kappa\mu_{0}}=0$ but $K_{+}\ket{\kappa\mu}\neq0$ for all $\mu\geq \mu_{0}$; therefore, the spectrum of the operator $K_0$ is the set $\{\mu=\mu_0+k:k=0,1,\ldots\}$.

Prior to continue, let us compute the coefficients $a_{\kappa\mu}$ and $b_{\kappa\mu}$ involved in the action of the generators on the basis. To this end we compute
$$
C\ket{\kappa\mu}=
\begin{cases}
[\kappa]_{q}[\kappa+1]_{q}\ket{\kappa\mu}, \\
([K_0]_{q}[K_0-1]_{q}-K_{+}K_{-})\ket{\kappa\mu}=([\mu]_{q}[\mu-1]_{q}-a_{\kappa\,\mu-1}b_{\kappa\mu})\ket{\kappa\mu},
\end{cases}
$$
which implies the relation $a_{\kappa\,\mu-1}b_{\kappa\mu}=[\mu]_{q}[\mu-1]_{q}-[\kappa]_{q}[\kappa+1]_{q}=[\mu+\kappa]_{q}[\mu-\kappa-1]_{q}$. On the other hand, taking into account the adjointness properties, we have
$$
\bra{\kappa\,\mu+1}K_{+}\ket{\kappa\mu}=
\begin{cases}
a_{\kappa\mu}, \\
\bra{\kappa\mu}K_{-}\ket{\kappa\,\mu+1}=b_{\kappa\,\mu-1},
\end{cases}
$$
so $a_{\kappa\mu}=b_{\kappa\,\mu-1}$, which leads to
$$
a_{\kappa\mu}=\sqrt{[\mu-\kappa]_{q}[\mu+\kappa+1]_{q}}
\quad\text{and}\quad
b_{\kappa\mu}=\sqrt{[\mu+\kappa]_{q}[\mu-\kappa-1]_{q}}.
$$
Therefore, the action of the infinitesimal generators on the basis is
$$
K_{+}\ket{\kappa\mu}=\sqrt{[\mu-\kappa]_{q}[\mu+\kappa+1]_{q}}\ket{\kappa\,\mu+1}
\quad\text{and}\quad
K_{-}\ket{\kappa\mu}=\sqrt{[\mu+\kappa]_{q}[\mu-\kappa-1]_{q}}\ket{\kappa\,\mu-1}.
$$

We now turn to the identification of the unitary irreducible representations.
To this end, taking into account that $K_{-}\ket{\kappa\mu_0}=0$, which is equivalent to the condition $[\mu_0+\kappa]_{q}[\mu_0-\kappa-1]_{q}=0$, we arrive at two (apparently different, but closely related) situations. The first one corresponds to the condition $[\mu_0+\kappa]_{q}=0$, which leads to the two requirements: $\mu_{0}=-\re(\kappa)$ and $\im(\kappa)=\pi k_{0}/\gamma$, for some $k_{0}\in\mathbb{Z}$. Moreover, since $K_{+}\ket{\kappa\mu}\neq0$ for all $\mu\geq\mu_0$, we must require also $\re(\kappa)<0$. Therefore, we obtain 
$$
\re(\kappa)<0,\quad \im(\kappa)=\frac{\pi k_0}{\gamma}, \quad\text{and}\quad \mu=-\re(\kappa),-\re(\kappa)+1,\ldots
$$
The second situation arises when the condition $[\mu_0 - \kappa - 1] = 0$ is satisfied, which implies that $\mu_0 = \re(\kappa) + 1$ and $\im(\kappa) = \pi k_0 / \gamma$, for some $k_0 \in \mathbb{Z}$, not necessarily equal to the one related to the previous situation. 
This time the constraint $K_{+}\ket{\kappa\mu}\neq0$ for all $\mu\geq\mu_0$ leads to the restriction $\re(\kappa)>-1$. Thus, in this case, we have
$$
\re(\kappa)>-1, \qquad \im(\kappa)=\frac{\pi k_0}{\gamma}, \quad\text{and}\quad \mu=\re(\kappa)+1,\re(\kappa)+2,\ldots
$$

Both kinds of representations are essentially the same, since replacing $\kappa$ with $-(\kappa + 1)$ interchanges their roles. As indicated at the beginning of the appendix, the eigenvalues remain invariant under this transformation. Furthermore, as can be observed, the possible values of $\kappa$ are not constrained by the algebraic structure itself, but rather by the physical phenomena to which they are applied.
This perspective is also highlighted in \cite[p.~1939]{aiz}, although in that work the eigenvalue $[\kappa]_q[\kappa - 1]_q$ is taken for the Casimir operator. This is consistent with our results here, through the substitution of $\kappa$ by $\kappa + 1$ in their formulation. However, it is customary in most mathematical physics papers (despite this awareness and regardless of potential applications) to adopt as a basis for a unitary and irreducible representation of the positive discrete series the following:
$$
\ket{\kappa\mu}, \qquad \kappa=0,\frac{1}{2},1,\frac{3}{2},\ldots, \qquad \mu=\kappa+1,\kappa+2,\ldots,
$$
which is usually denoted by~$D^{\kappa+}$.

\section{The $q$-Hahn and $q$-dual Hahn polynomials}\label{AA}
For the sake of completeness, we include here some properties of the hypergeometric polynomials on non-uniform lattices and, in particular of the so-called $q$-Hahn polynomials. More details of these mathematical objects can be found, e.g., in~\cite{ran-q,nu,nsu}.

The hypergeometric polynomials on non-uniform lattices are the polynomial solutions, $y_n(s):=y_n(x(s))=a_n [x(s)]^n+\cdots$, $a_n\neq0$, of the second order linear difference $q$-Nikiforov-Uvarov equation  
$$
\sigma(s)  \frac{\Delta}{\Delta x(s-\mbox{\footnotesize 1/2})} 
\frac{\nabla y(s)}{\nabla x(s)} + \tau(s)  \frac{\Delta y(s)}{\Delta x(s)}
+ \lambda_n y(s) =0,\quad x(s)=c_1[q^{s}+q^{-s-\mu}]+ c_3,
$$
where $c_1,c_2,c_3,\mu\in\mathbb{C}$,  $\sigma(s)=q^{-2s}(q^s-q^{s_1})(q^s-q^{s_2})(q^s-q^{s_3})(q^s-q^{s_4})$, $\tau(s)=\tau' x(s)+\tau_0$, $\lambda_n$ are the corresponding eigenvalues, $a_n$ are the leading coefficients, and $\nabla f(s)=f(s)-f(s-1)$, $\Delta f(s)=f(s+1)-f(s)$, denote the backward and forward difference operators, respectively. For our purposes it is convenient to write the above equation it the following equivalent form:
\begin{equation}\label{dif-eq-pol}
A(s)y(s+1) + B(s)y(s) + C(s)y(s-1) + \lambda _n y(s) = 0,\quad
B(s)=-A(s)-C(s),
\end{equation}
where 
$$
\phi(s)=\sigma (s) + \tau (s) \Delta x(s-\mbox{\footnotesize 1/2}),\
A(s)=\frac{\phi(s)}{\Delta x(s) \Delta x(s-\mbox{\footnotesize 1/2})}, \text{ and }
C(s)=\frac{\sigma (s) }{\nabla x(s) \Delta x(s-\mbox{\footnotesize 1/2})}.
$$
They satisfy the orthogonality relation (it is assumed that $b-a=N\in\mathbb{N}$) 
\begin{equation}\label{ort-rel-pol}
\sum\limits_{s = a}^{b - 1} y_n (s) y_m (s) \rho (s)\Delta
x(s -\mbox{\footnotesize 1/2}) =
\delta _{nm} d_n^2,\quad s=a,a+1,\dots, a+N=b,
\end{equation}
where $\rho$ and $d_n$, are the corresponding weight function and norm, respectively. A consequence of this orthogonality relation is the  following so-called three-term recurrence relation: 
\begin{equation}\label{ttrr-pol}
x(s)y_n (s) = \alpha _n y_{n + 1} (s) + \beta _n y_n (s) + 
\gamma _n y_{n - 1} (s),\quad y_{-1}(s):=0,\quad  y_{0}(s):=1.
\end{equation}
The polynomials $y_n(s)$ also satisfy the so-called lowering and rising relations 
\begin{small}\begin{align}\label{low-rel}
\sigma(s) \frac{\nabla  y_{n}(s) }{\nabla x(s)} =  -
\frac{\lambda_n}{\lambda_{2n+1}} \frac{[2n+1]_q}{[n]_q}  & \bigg[ \tau_n(s) y_{n}(s)  -
\frac{B_n}{B_{n+1}} y_{n+1}(s)\bigg], \\ \label{rai-rel}
\phi(s) \frac{\Delta  y_{n}(s)}{\Delta x(s)} = -
\frac{\lambda_n}{\lambda_{2n+1}} \frac{[2n+1]_q}{[n]_q} &  \left[ 
\left(  \tau_n(s) + \frac{[n]_q \, \lambda_{2n+1} }{[2n+1]_q} \Delta x(s-\mbox{\footnotesize$\frac12$})\right) y_{n}(s)-\frac{B_n}{B_{n+1}}y_{n+1}(s)\right],
\end{align}\end{small}
respectively, where $B_n$ is a normalizing factor, and
$$
\tau_n(s)=\frac{\phi(s+n)-\sigma(s)}{ \Delta x(s+\mbox{\footnotesize$\frac{n-1}2$})}.
$$

The monic $q$-Hahn polynomials on the lattice $x(s)=(q^{2s}-1)/(q^2-1)$ were introduced in \cite[Eq.~(3.11.53), p.~150]{nsu} and detailed studied in \cite{arv}. They are given by
\begin{equation}
\label{q-hahn1-hyp}
h_n^{\alpha,\beta}(s,N)_q  =
(-1)^n q^{{n}(\alpha+\beta+1+\frac{n+1}{2})}\frac{(\beta\!+\!1|q)_n[N-1]_q!}{[N-n-1]_q![n]_q!}
{_3}F_2\bigg(\!\begin{array}{c}
-n\, , \,  -s \, , \,  \alpha+\beta+n+1   \\
\beta+1 \, , \, 1 -N 
\end{array}\!\bigg\vert\, q\, , \, q^{s-N-\alpha}\!\bigg),
\end{equation}
The monic dual $q$-Hanh polynomials $W^{(c)}_{n}(s,a,b)$ on the lattice $x(s)=[s]_q[s+1]_q$ were introduced in \cite[Eq.~(3.11.36), p.~146]{nsu} and detailed studied in 
\cite{ran-dual,ran-q} and are given by
\begin{equation}
\label{q-hahn2-hyp}
W_{n}^{(c)}(s,a,b)_q=\frac{(a-b+1|q)_{n}(a+c+1|q)_n}{q^{n(b-c-1+(n-1)/2)}\qn{n}!}
{_3}F_2\bigg(\!\begin{array}{c}
-n\, , \,  a-s \, , \,  a+s+1   \\
a-b+1 \, , \, a+c+1 
\end{array}\!\bigg\vert\, q\, , \, q^{b-c-n}\!\bigg).
\end{equation}

The $q$-Hahn and dual $q$-Hanh polynomials satisfy the orthogonality relation \eqref{ort-rel-pol} where $\rho$ and $d_n^2$ are given by 
\begin{align*}
	\rho(s) & =  \displaystyle q^{( {\alpha+\beta})s}\frac{[s+\beta]_q!
		[N+\alpha-s-1]_q!}{[s]_q![N-s-1]_q!},\quad \alpha,\beta>-1,\quad a=0,\, b=N,\, N \in\mathbb{N},\\
	d_n^{2} & =\displaystyle \frac{q^{N({\alpha+\beta}+{2}N)} \, q^{-\frac{\alpha(\alpha-3)+2}{2}}
		[n+\alpha]_q![n+\beta]_q![n+\alpha+\beta+N]_q!}{q^{n\left({\alpha-\beta}+{2}N+2\right)}\qn{n}!\qn{N-n-1}!\qn{n+\alpha+\beta}!\qn{2n+\alpha+\beta+1}!},
\end{align*} 
and 
\begin{align*}
\rho(s) & =  \frac{q^{-s(s+1)} [s+a]_q! [s+c]_q! }
{[s-a]_q!  [s-c]_q! [s+b]_q! [b-s-1]_q!}, \quad 
a>-1/2,\, b=a+N,\, N \in\mathbb{N},\, |c|<1+a,\\
d_n^2 & =  \frac{q^{(ac-ab-bc+a+c-b+1+ 2n(a+c-b)-n^2+5n)}[a+c+n]_q!}
{[n]_q!  [b-c-n-1 ]_q!  [b-a-n-1]_q! },
\end{align*}
respectively. The coefficients of the difference equation \eqref{dif-eq-pol} and the three-term recurrence relation \eqref{ttrr-pol}, as well as the normatization factor $B_n$ and $\tau_n$ can be found in Table \ref{tabla-hahn}.

Let us point out that we have written the expressions for $\rho$ and $d_n^2$ in terms of the $q$-factorials since they are evaluated in non-negative integers. However, in the general case, one should use the symmetric $q$-Gamma function defined in~\cite[Eq.~(3.2.24), p.~67]{nsu}.

 \begin{table}[ht!]
\caption{\label{tabla-hahn}Main data for the $q$-Hahn polynomials ($n=0,1,\dots,N-1$).} 
{\small \begin{center}{\renewcommand{\arraystretch}{2}
\begin{tabular}{@{}|c|c|c|}\hline   
&    $h_n^{\alpha,\beta}(s,N)_q$, \, $x(s)=\frac{q^{2s}-1}{q^2-1}$  & 
$W_n^{(c)}(s,a,b)_q$,\, $x(s)=[s]_q [s+1]_q$
\\
\hline\hline
$\sigma(s)$ & $-q^{2s-2}[s]_q[N+\alpha-s]_q$ &   $q^{s+c+a-b+2} [s-a]_q[s+b]_q[s-c]_q$
\\ \hline  
$\phi(s) $ & $-q^{2s+{\alpha+\beta}}[s+\beta+1]_q[s-N+1]_q$ &   
$q^{-s+c+a-b+1} [s+a+1]_q[b-s-1]_q[s+c+1]_q$
\\ \hline  
$\lambda_n$ & $q^{{\beta+2-N}}[n]_q[n+\alpha+\beta+1]_q$ &  $q^{-n+1}[n]_q$
\\ \hline  
$\alpha _n$ & $  \frac{q^{-{\beta+2-N}}[n+1]_q[n+\alpha+\beta+1]_q}{[2n+\alpha+\beta+2]_q[2n+\alpha
+\beta+1]_q}$ & $q^{3n} [n+1]_q$\\ 
\hline $\beta_n $ & $  q^{2\alpha+2N+{n}-2}\frac{[n+\alpha+\beta+1]_q[n+\beta+1]_q[N-n-2]_q}{[2n+\alpha
+\beta+2]_q[2n+\alpha+\beta+1]_q}$ &
$q^{2n-b+c+1} [b-a-n+1]_q [a+c+n+1]_q +$ \\
& $ + \frac{q^{-(2N+\beta+n+3(\alpha+1))}[n+\alpha]_q[n+\alpha
+\beta+N]_q[N-n]_q[n]_q}{[2n+\alpha+\beta+1]_q[2n+\alpha+\beta]_q^{2}[2n
+\alpha+\beta-1]_q[N-n-1]_q}$ & 
$+q^{2n+2a+c-b+1} [n]_q [b-c-n]_q + [a]_q [a+1]_q $
\\ \hline 
$\gamma_n$ & $  \frac{q^{-N-\alpha-4}[n
+\alpha]_q[n+\beta]_q[n+\alpha+\beta+N]_q[N-n]_q}{[2n+\alpha+\beta
+1]_q[2n+\alpha+\beta]_q^{2}[2n+\alpha+\beta-1]_q} $ &  
$q^{n+3+2(c+a-b)} [n+a+c]_q[b-a-n]_q[b-c-n]_q $
\\ \hline  
$B_n$ & $\frac{(-1)^n}{[n]_q!}$ &   $\frac{(-1)^n}{[n]_q!}$ 
\\
\hline
$\tau_n$ & $ q^{n+\alpha+\beta+1}[s+n+\beta+1]_q[N-s-n-1]$  &  $-q^{2n}[s+\khalf]_q[s+\khalf+1]_q   
+ q^{c-b+n+1} [c+\khalf]_q[b-\khalf]_q $ \\
& $+q^{-n-1}[s]_q[N+\alpha-s]_q$  & $ +q^{a+c-b+1-\khalf} 
[a+\khalf+1]_q[b-c-n-1]_q $  \\
\hline
\end{tabular}
}\end{center} }
 \end{table}

\begin{landscape}
\section{Some linear recurrence relations of the Clebsch-Gordan coefficients for mixed representations}\label{BB}
The present appendix is devoted to the formulation of some linear recurrence relations of the Clebsch-Gordan coefficients for mixed representations.
\small

\begin{multline*}
c_1\sqrt{[\mu'-\kappa']_q[\mu'+\kappa']_q}\braket{\kappa\mu,jm}{\kappa'\mu'}
=[\kappa+m+\kappa'+1]_q\sqrt{[\mu'+\kappa'+1]_q[j+m+1]_q[\mu'+\kappa']_q}q^{2(j-\kappa)}\braket{\kappa\mu,j\,m+1}{\kappa'\,\mu'+1} \\
+[\kappa-\kappa'-m+1]_q\sqrt{[\mu'-\kappa'-1]_q[j-m+1]_q[\mu'-\kappa']_q}q^{2(\kappa+1)}\braket{\kappa\mu,j\,m-1}{\kappa'\,\mu'-1},
\end{multline*}
where
$$
c_1=[\kappa'-\kappa+j]_q[\kappa'+\kappa-j+1]_{q}q^{-(3j+\kappa'+2m+3)}-[j-m]_q[\kappa+m+\kappa'+1]_{q}q^{-\kappa+1}+[j+m]_q[\kappa-\kappa'-m+1]_{q}q^{\kappa+1}.
$$

\vspace{0.5cm}

\begin{multline*}
c_2\sqrt{[2\kappa'-1]_q[\mu'-\kappa'-1]_q}\braket{\kappa\mu,jm}{\kappa'\mu'} \\
=\sqrt{[k-j+\kappa']_q[\kappa+j-\kappa'+1]_q[\mu'-\kappa']_q[\mu'-\kappa'-1]_q[\kappa'-\kappa-+j]_q[\kappa'+\kappa+j+1]_q[\kappa'+\mu']_q}q^{\mu'-2j-1}\braket{\kappa\mu,jm}{\kappa'-1\,\mu'} \\
-[2\kappa']_q\sqrt{[j+m+1]_q[j-m]_q[\kappa'+\mu'+1]_q[2\kappa'-1]_q[\mu'-\kappa'-1]_q}q^{-2j-2m-\mu}\braket{\kappa\mu,j\,m+1}{\kappa'\,\mu'+1},
\end{multline*}
where
\begin{multline*}
c_2=[\kappa'-\kappa+j]_q[\kappa'+\kappa+j+1]_q[\kappa'+\mu']_{q}q^{\kappa'-\mu'-2j+1}+[2\kappa']_q\left(\left[\kappa+j+\frac{j+m}{2}+1\right]_q[\kappa+\mu]_{q}q^{-2j-1-\mu'+(j+\mu)/2}\right. \\
\left.+\left[\mu-\frac{j+m}{2}\right]_q\left[\kappa+j+1-\frac{j+m}{2}\right]_{q}q^{-(\kappa+\mu)}-\left[\kappa'+\frac{j+m}{2}\right]_q\left[\kappa'+\frac{j+m}{2}+1\right]_{q}q^{2(j+m)}\right)q^{j+m+1}.
\end{multline*}

\vspace{0.5cm}

\begin{multline*}
c_3\sqrt{[2\kappa'+3]_q[2\kappa'-1]_q}\braket{\kappa\mu,jm}{\kappa'\mu'} \\
=\sqrt{[2\kappa+1]_q[2\kappa'-1]_q[\kappa+j+\kappa'+2]_q[j-\kappa+\kappa'+1]_q[\mu'+\kappa'+1]_q[\kappa'+\kappa-j+1]_q[\kappa+j-\kappa']_q[\mu'-\kappa'-1]_q}\braket{\kappa\mu,jm}{\kappa'+1\,\mu'} \\
-\sqrt{[2\kappa'+1]_q[2\kappa'+3]_q[\kappa'-\kappa+j]_q[\kappa'+\kappa+j+1]_q[\mu'+\kappa']_q[\kappa-j+\kappa']_q[\kappa+j-\kappa'+1]_q[\mu'-\kappa']_q}\braket{\kappa\mu,jm}{\kappa'-1\,\mu},
\end{multline*}
where
$$
c_{3}=[j+m]_{q}q^{\mu'+j-m}+[\kappa'+\kappa-j+1]_q[\kappa+j-\kappa']_q[\mu'-\kappa'-1]_{q}q^{-\kappa'}-[\kappa'-\kappa+j]_q[\kappa'+\kappa+j+1]_q[\kappa'+\mu']_{q}q^{\kappa'-2}.
$$

\end{landscape}

\end{appendices}


\end{document}